\documentclass{aa} 
\usepackage{txfonts} 
\usepackage{graphicx,amssymb,subfigure}
\usepackage[normalem]{ulem}

\begin{document}

%
%
\newcommand{\wav}[1]{$\lambda#1\,\mathrm{cm}$}  
\newcommand{\wwav}[2]{$\lambda\lambda#1,#2\,\mathrm{cm}$}  
\newcommand{\wwwav}[3]{$\lambda\lambda#1,#2,#3\,\mathrm{cm}$}  
\newcommand{\dif}{\,\mathrm{d}} 
\newcommand{\HII}{\mathrm{H\,\scriptstyle II}} 
\newcommand{\Brpe}{B_{\perp}}
\newcommand{\Brpa}{B_{\parallel}}
\newcommand{\pco}{p_\mathrm{CO}}
\newcommand{\pii}{p_\mathrm{I}} 
\newcommand{\piso}{p_\mathrm{ISO}}
\newcommand{\ppi}{p_\mathrm{PI}}
\newcommand{\pB}{p_\mathrm{B}} 
%
%
\newcommand{\cm}{\,\mathrm{cm}} 
\newcommand{\mm}{\,\mathrm{mm}}
\newcommand{\cmcube}{\,\mathrm{cm^{-3}}} 
\newcommand{\dyn}{\,\mathrm{dyn}}
\newcommand{\erg}{\,\mathrm{erg}} 
\newcommand{\g}{\,\mathrm{g}} 
\newcommand{\Hz}{\,\mathrm{Hz}}
\newcommand{\GHz}{\,\mathrm{GHz}} 
\newcommand{\Jy}{\,\mathrm{Jy}} 
\newcommand{\Jyb}{\,\mathrm{Jy/beam}} 
\newcommand{\cms}{\,\mathrm{cm\,s^{-2}}}
\newcommand{\kms}{\,\mathrm{km\,s^{-1}}} 
\newcommand{\mJy}{\,\mathrm{mJy}} 
\newcommand{\mJyb}{\,\mathrm{mJy/beam}} 
\newcommand{\K}{\,\mathrm{K}} 
\newcommand{\kpc}{\,\mathrm{kpc}}
\newcommand{\Mpc}{\,\mathrm{Mpc}} 
\newcommand{\mG}{\,\mathrm{mG}} 
\newcommand{\MHz}{\, \mathrm{MHz}} 
\newcommand{\Msol}{\,\mathrm{M_\sun}}
\newcommand{\mum}{\,\mu\mathrm{m}} 
\newcommand{\n}{\,n_\mathrm{e}} 
\newcommand{\pc}{\,\mathrm{pc}}
\newcommand{\s}{\,\mathrm{s}} 
\newcommand{\uG}{\,\mu\mathrm{G}} 
\newcommand{\uJy}{\,\mu\mathrm{Jy}} 
\newcommand{\uJyb}{\,\mu\mathrm{Jy/beam}} 
\newcommand{\yr}{\,\mathrm{yr}}

\title{Analysis of spiral arms using anisotropic wavelets: gas, dust and
magnetic fields in M51}

\titlerunning{Gas, dust and magnetic spiral arms in M51}
\authorrunning{Patrikeev et al.}

\author{I.~Patrikeev \inst{1} 
	\and A.~Fletcher \inst{2,3} 
	\and R.~Stepanov \inst{1} 
	\and R.~Beck \inst{2} 
	\and E.~M.~Berkhuijsen \inst{2} 
	\and P.~Frick \inst{1} 
	\and C.~Horellou \inst{4}}

\institute{Institute of Continuous Media Mechanics, Korolyov str.~1, 614061
Perm, Russia 
\and Max-Planck-Institut f\"ur Radioastronomie, Auf dem H\"ugel 69,
53121 Bonn, Germany 
\and School of Mathematics and Statistics, University of
Newcastle, Newcastle upon Tyne NE1 7RU, UK 
\and Onsala Space Observatory,
Chalmers University of Technology, 439 92 Onsala, Sweden}

\offprints{I.~Patrikeev. Current address: 301 University Blvd., UTMB, Galveston,
TX, USA, 77550-0456, \email{igpatrik@utmb.edu}}

\date{Received / Accepted }

\abstract{The origin of the spiral pattern of magnetic fields in disc galaxies
is an open question. Comparison of the regular magnetic field orientation with
the gaseous spiral arm pitch angles can tell us whether spiral shock compression
is responsible for the magnetic spirals. We also wish to see whether the ridges
of different components of the ISM show the large-scale, systematic shifts
expected from density wave theory. We have developed a technique of isolating
elongated structures in galactic images, such as spiral arms, using anisotropic
wavelets and apply this to maps of the CO, infrared and radio continuum emission
of the grand-design spiral galaxy M51. Systematic shifts between the ridges of
CO, infrared and radio continuum emission that are several $\kpc$ long are
identified, as well as large variations in pitch angle along spiral arms, of a
few tens of degrees. We find two types of arms of polarized radio emission: one
has a ridge close to the ridge of CO, with similar pitch angles for the CO and
polarization spirals and the regular magnetic field; the other does not always
coincide with the CO arm and its pitch angle differs from the orientation of its
regular magnetic field. The offsets between ridges of regular magnetic field,
dense gas and warm dust are compatible with the sequence expected from spiral
density wave triggered star formation, with a delay of a few tens of millions of
years between gas entering the shock and the formation of giant molecular clouds
and a similar interval between the formation of the clouds and the emergence of
young star clusters. At the position of the CO arms the orientation of the
regular magnetic field is the same as the pitch angle of the spiral arm, but
away from the gaseous arms the orientation of the regular field varies
significantly. Spiral shock compression can explain the generation of one type
of arm of strong polarized radio emission but a different mechanism is probably
responsible for a second type of polarization arm.

 }

\maketitle

\section{Introduction} 
\label{sec:intro}

Disc galaxies often display spiral patterns in their distributions of stars,
gas, dust and magnetic fields. Spiral structure is present in both the
\emph{distribution} of the radio synchrotron intensity, and the
\emph{orientation} of the $\kpc$-scale regular field, as measured by the
$B$-vectors of radio polarization (e.g. Beck\ \cite{Beck05}). Comparing the
location and orientation of the spiral patterns in different components of the
interstellar medium (ISM) can provide important information on the
astrophysical connections between interstellar gas, dust and magnetic fields in
galaxies.

Usually the prominent arms in disc galaxies are treated as having a logarithmic
spiral pattern, where the pitch angle of the arm is constant along its length.
However, Kennicutt (\cite{Kennicutt81}) found that the logarithmic spiral form
is no better a mathematical description of real spirals than hyperbolic forms,
and that ``\dots small scale distortions preclude the possibility of \emph{any}
universal shape for galactic spirals''. Deviations from the logarithmic spiral
may be caused by local disturbances or arise from systematic global effects,
such as the gravitational force of a nearby galaxy or the presence of more than
one spiral density wave (Elmegreen et al.~\cite{Elmegreen89}). If
the regular magnetic field spirals are being aligned with the gaseous spiral
arms by shock compression, the magnetic field orientation should be closely
related to the local spiral arm pitch angle. In order to quantify how well the
different spirals are aligned a robust method of measuring local pitch angles is
required.

M51 is probably perturbed by a recent encounter with its companion galaxy
NGC~5195. Such interactions often result in increased star formation rates,
either localised or global, as tidal forces and spiral density waves compress the
interstellar medium. The two spiral arms of M51 can be traced through more than
360\degr\ in azimuthal angle in numerous wavebands and several authors have
investigated their structure. Elmegreen et al. (\cite{Elmegreen89}) de-projected
images taken in the optical B and I bands into the log(radius)--azimuth plane
and used the observed amplitude variations along the arms to locate resonances
of spiral modes. They concluded that M51 contains an inner and an outer system
of spiral arms with a conjunction of resonances at 2\arcmin\,-- 3\arcmin\
galacto-centric radius, triggered by the companion. Measurements of the pitch
angles of the main dust lanes by Howard \& Byrd (\cite{Howard90}) and of the
variation in star formation efficiency along the arms by Knapen et al.
(\cite{Knapen92}) confirmed this double spiral structure. It is clear from a
cursory visual inspection that even the well defined --- grand design --- spiral
arms of M51 have many variations in pitch angle along their length.

Radial shifts between the locations of the crest of the spiral arms in different
constituents are expected from density wave theory. Petit et al.
(\cite{Petit96}) observed a 300-400~pc shift between H$\alpha$ and UV arms. Rand
\& Kulkarni (\cite{Rand90}), Scoville et al. (\cite{Scoville01}), and Tosaki et
al. (\cite{Tosaki02}) noted that the H$\alpha$ arms are located outside the CO
arms. Henry et al. (\cite{Henry03}) found that the Paschen-$\alpha$ line
emission is offset from the CO emission along part of the arms. All these shifts
refer to the inner spiral system and are expected for arms triggered by density
waves. Tilanus et al. (\cite{Tilanus88}) found that thermal radio continuum and
H$\alpha$ arms are significantly shifted outwards from the dust lanes, whereas
the non-thermal radio continuum arms occur just inside the dust lanes.

Most of the above analyses were made by overlaying the contours of one image on
a grey-scale map of the other or by making azimuthal cuts through the data at
different radii. Both methods, while showing the most significant local
displacements in the position of the arms, make it difficult to see the
extent along the arm of any systematic shift. In the present paper we describe an
objective method, using wavelet analysis, to determine the location of the
spiral arm ridges, and measure their local pitch angles, along the entire length
of an arm. We apply this method to images of the total and polarized radio
continuum emission at \wav{6} (Fletcher et al. in preparation), the dust
emission at 15$\mum$ (Sauvage et al.~\cite{Sauvage96}) and the CO(1--0) line
emission (Helfer et al.~\cite{Helfer03}).

In Section~\ref{sec:method} we describe the anisotropic wavelet transform and
how we use this method to identify the location of a spiral arm and to measure
its localised pitch angles. The observational data we use are briefly discussed
in Section~\ref{sec:data} and the results of applying the anisotropic wavelet
method to the data are given in Section~\ref{sec:results}. The astrophysical
implications of the results are discussed in Section~\ref{sec:discuss}.

\section{The Method} 
\label{sec:method}

In image analysis, wavelet based methods allow the isolation of features, such
as spiral arms, and the decomposition of a map into a hierarchy of structures on
different scales. Wavelets are a tool for data analysis based on self-similar
functions which are well localised both in the physical and frequency domains.
Using one-dimensional isotropic wavelets Frick et al.~(\cite{Frick00})
identified systematic shifts between the magnetic and optical spiral arms in
NGC~6946 and using two-dimensional isotropic wavelets Frick et
al.~(\cite{Frick01}) investigated the scale-by-scale correlations between maps
of the same galaxy in different spectral ranges. In this paper a new method,
using a two-dimensional \emph{anisotropic} wavelet is presented.

The wavelet transform can be considered as a generalization of the Fourier
transform. The classical Fourier transform is based on harmonic functions. The
generalized Fourier transform allows using non-harmonic orthogonal functions as
a basis (e.g. a Walch set of discrete, piecewise-constant functions). The
short-time Fourier transform and the Gabor transform use oscillatory basis
functions with local support. The wavelet transform also uses oscillatory
functions, but in contrast to the classical Fourier transform these functions
decay rapidly toward infinity and all functions in the wavelet basis are
self-similar (the main distinction between the wavelet transform and the Gabor
transform). One-dimensional and isotropic multi-dimensional wavelet transforms
are based on the space-scale decomposition of the signal (in other words, the
family of wavelets has two parameters, governing the location and the size of
the basis function). Using the continuous isotropic wavelet transform, a 2D
image is decomposed into a 3D cube of wavelet coefficients. Cross-sections of
the cube are slices which contain the image details at a fixed scale. As a
result, the wavelet transform conserves the local properties of the image at all
scales. If required, the original image can be reconstructed from the cube by
summing over all scales (this procedure is called the inverse wavelet
transformation).

An \emph{anisotropic} wavelet transform is the convolution of the image with a
set of wavelets having different locations, sizes and \emph{orientations}. Such
a family of basis functions is generated by translations, dilations and
\emph{rotations} of the basic wavelet. Applying the two-dimensional anisotropic
wavelet transform to an image generates a four-dimensional data set, which is a
space-scale-orientation decomposition. Fixing the space and scale parameters ---
based on some objective criteria --- enables one to track the orientation of an
elongated structure (Antoine~\cite{Antoine93}). An extended description of the
continuous wavelet transform can be found, for example, in Holschneider
(\cite{Holschneider95}) and Torresani (\cite{Torresani95}).

\subsection{The Texan Hat function}

The ideal anisotropic wavelet for astronomical image processing should combine
high angular sensitivity with a simple computational formula; the latter
requirement is due to the time required to calculate the wavelet transform of a
high resolution image. In this work we use a specially designed \emph{ad hoc}
anisotropic wavelet. Its formula is very simple and the wavelet transformation
can be calculated efficiently. We do not use the Cauchy wavelet
(Antoine~\cite{Antoine93}), which provides the best angular sensitivity, due to
its more complicated formula.

We introduce our wavelet by starting with the Mexican Hat function (MH), one of
the most commonly used wavelets. In the 1D case, the MH is described by
$\psi(x)=(1-x^2)\exp(-x^2/2)$. The 1D MH can be extended into 2D in one of two
ways: (i) constructing an isotropic (axisymmetric) function by rotation
of the 1D MH, (ii) translation of the 1D MH along an axis but
restricting it within a (e.g. Gaussian-shaped) window in this direction, to
obtain an anisotropic function. We call the anisotropic wavelet constructed
using the latter method the Texan Hat function (TH) and it can be represented by
\begin{equation}
\mathrm{\psi}(x,y,a) = \left[1-\left(\frac{y}{a}\right)^2 \right]
    \exp\left( -\frac{x^2+y^2}{2a^2}\right),
\label{eq:th}
\end{equation}
where $a$ is the scale parameter by which the wavelet is dilated. In the form of
Eq.~(\ref{eq:th}) the TH is sensitive to structures elongated along the x-axis.
This means that if an image is convolved with the TH, extended objects parallel
to the x-axis will be amplified. Substitution of $(x\cos\varphi+y\sin\varphi)$
for $x$ and $(y\cos\varphi-x\sin\varphi)$ for $y$ in Eq.~(\ref{eq:th}) makes the TH
sensitive to its orientation $\varphi$ measured from the x-axis, since this
substitution is equivalent to a rotation of the axes by $\varphi$.  A complete
set of basis functions of different sizes and orientations is obtained by
dilation and rotation of the basic wavelet. Note that since the TH is a
symmetric function with respect to rotation by 180\degr\, it is sensitive to the
\emph{orientation} of an elongated structure, not to its \emph{direction}:
\begin{equation}
\mathrm{\psi}_{\varphi}(x,y) = \mathrm{\psi}_{\varphi + 180^{\circ}}(x,y).
\label{eq:symm}
\end{equation}

Another, more formal, way to design the TH is as follows. The 1D MH can be
defined as the second derivative of the Gaussian function. Similarly, the TH is
defined as the second partial derivative of the 2D Gaussian function:
\begin{equation}
\mathrm{\psi}(x,y) = -\frac{\partial ^2}{\partial y^2}
    \exp\left( -{{x^2+y^2}\over{2}} \right).
\label{eq:deriv}
\end{equation}
Equation~(\ref{eq:deriv}) can be parametrized for variation in its size $a$ by
substituting $x$ with $x/a$ and $y$ with $y/a$ (i.e. a dilation of the axes).
This approach is useful for determining the properties of the TH in the Fourier
domain. For example, one can easily estimate the Fourier transform of the TH by
using the Fourier transform of the Gaussian function and the theorem of
derivatives in the Fourier domain. Furthermore the definition of the wavelet as
a derivative is not only valid for the 2D case, but also for higher dimensions.

The anisotropy of the TH can be successfully exploited for an analysis of the
orientations of galactic spiral arms. A brief overview of the method was first
presented in Patrikeev et al.~(\cite{Patrikeev05}). A detailed description of
the technique and an illustrative application to an artificial spiral image are
given in the next section.

\subsection{Measurement of spiral arm position and pitch angles}

\begin{figure}[htbp]
\includegraphics[width=0.24\textwidth]{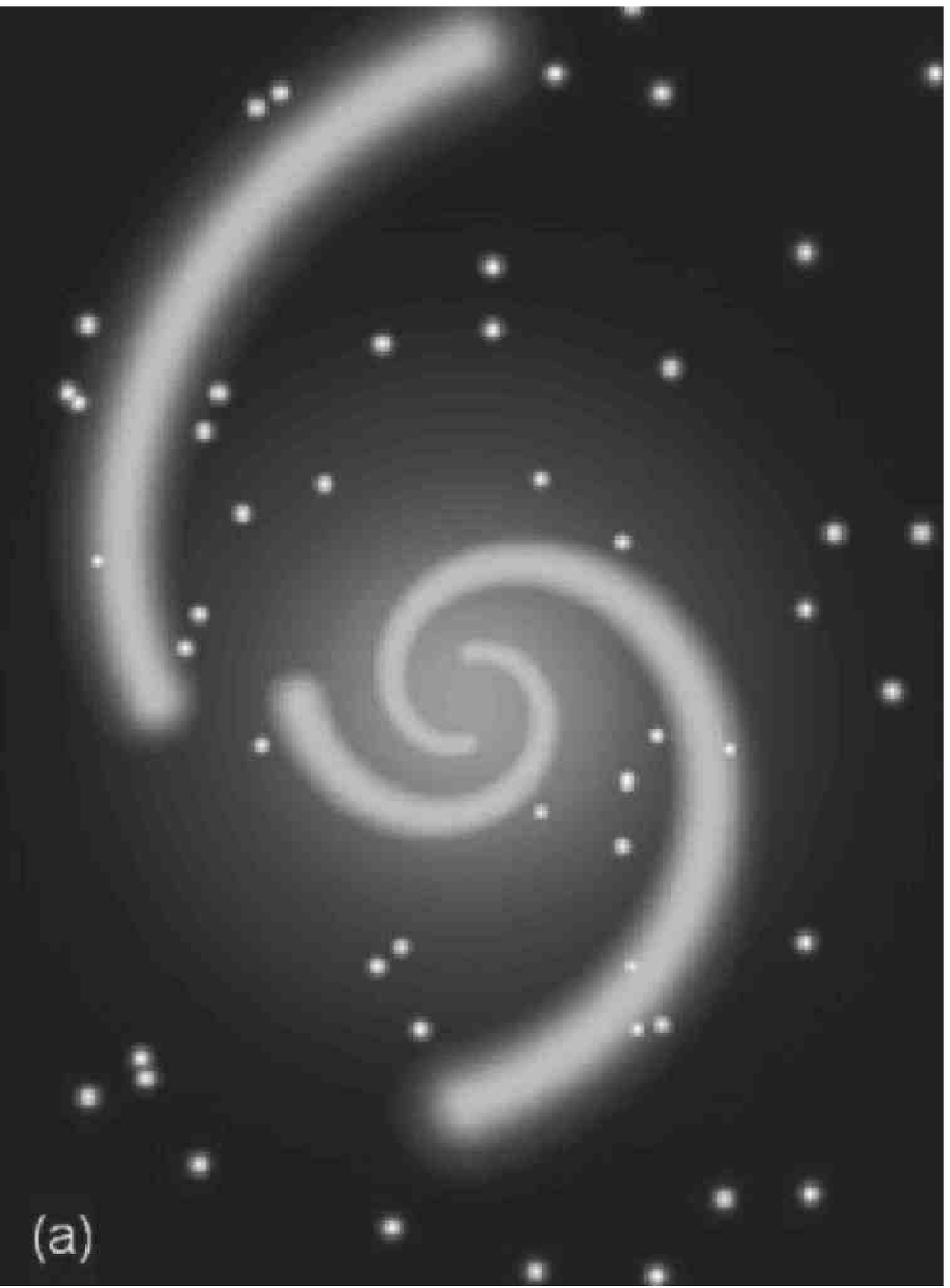}
\includegraphics[width=0.24\textwidth]{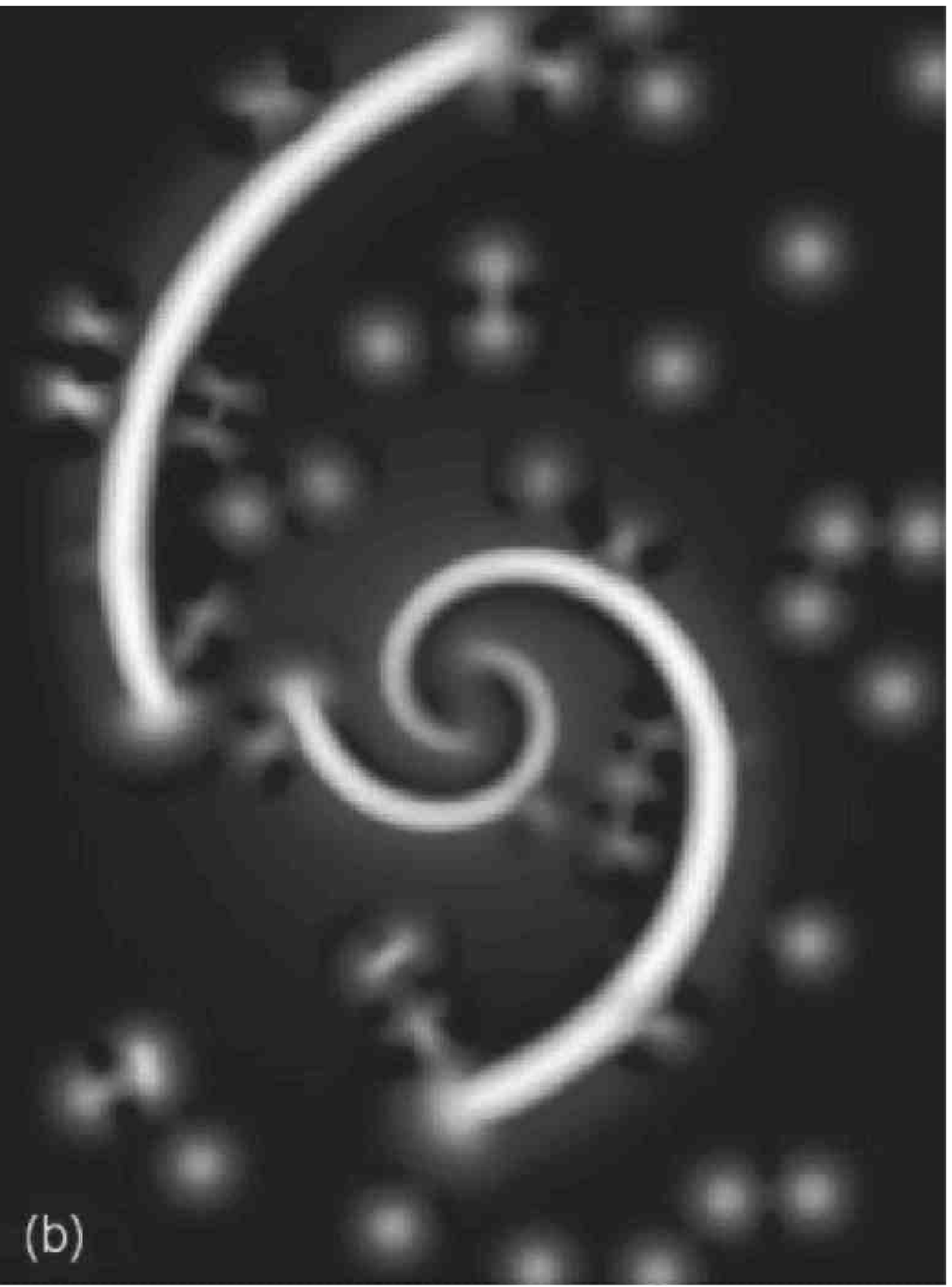}
\caption{Analysis of an artificial spiral image with the Texan Hat function.
\textbf{(a)} Two arm logarithmic spiral with a central exponential disc and
randomly distributed, slightly smoothed, point sources. The pitch angle of the
first (unbroken) arm is 20\degr; the pitch angle of the second (broken) arm is
15\degr\ at small radii and 25\degr\ at large radii. \textbf{(b)} Map of the
maximum wavelet coefficients at a fixed scale that is close to the average arm
width.}
\label{fig:test}
\end{figure}

\begin{figure}[htbp]
\includegraphics[width=0.48\textwidth]{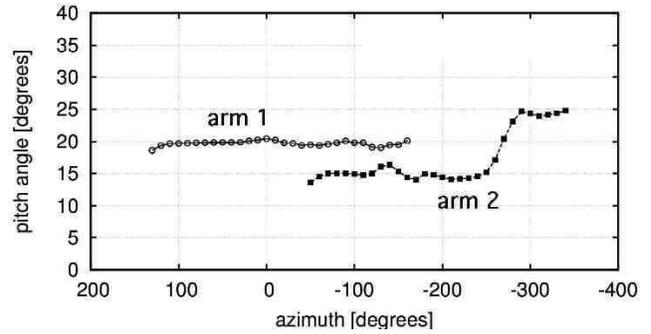}
\caption{Azimuthal variation of pitch angle determined from the anisotropic
wavelet transform of the spiral arms of Fig.\ref{fig:test}a. Pitch angles are
measured at fixed azimuthal increments along the ridges of the arms. The
azimuthal angle is measured counter-clockwise from the top of the image.}
\label{fig:test-res}
\end{figure}

It is obvious that for an isotropic structure (e.g. a round-shaped object, or
randomly distributed set of spots, etc.) an orientation cannot be defined. In
this work we use the TH wavelet to trace the ridge of a spiral arm and to
measure its local pitch angles. To illustrate how the TH can determine the
position and pitch angle of a spiral arm, we use the artificial image presented
in Fig.~\ref{fig:test}(a). The test image consists of: (1) two
logarithmic spiral arms with Gaussian cross-sectional profiles, the first has a
constant pitch angle of 20\degr\ and the second, broken, arm has a pitch angle of
15\degr\ at small radii and 25\degr\ at large radii; (2) an exponential disc
that is about $70\%$ of the arm intensity in the central region; (3) fifty
bright point sources, slightly smoothed with a Gaussian kernel, distributed
randomly over the image.

The test image was convolved with a TH at a fixed scaling parameter $a$ close to
the average spiral width and a rotation parameter $\varphi$ ranging from 0\degr\
to 180\degr\ in steps of 1\degr. From the convolution we obtained a stack of
maps of wavelet coefficients; each map represents a specific value of $\varphi$
and has the same size as the test image. Figure~\ref{fig:test}(b) shows the map
of maximum wavelet coefficients, obtained by taking the largest value through
the stack at each position. The central exponential background is invisible
because the wavelet coefficients of such a structure are very low. The point
sources are converted to faint spots; their wavelet coefficients have no
pronounced maximum since a round source has no orientation. In contrast to the
non-oriented structures, the two-fold spiral becomes more emphatic, with large
wavelet coefficients.

We define the location of the ridge of a spiral arm as the position where the
maximum wavelet coefficient is found. The local pitch angle is defined as the TH
orientation corresponding to the maximum wavelet coefficient.
Figure~\ref{fig:test-res} shows the measured azimuthal
variation of pitch angle at the ridge of the spirals 
(the spike-like deviations were suppressed). The first arm has a pitch
angle close to the expected value of 20\degr. The pitch angle of the second arm
changes from 15\degr\ to 25\degr\ (near the azimuth about -270\degr) remaining
constant before and after the break. Small deviations of the pitch angle are
caused by the influence of bright sources located in the vicinity of the
spirals.

The analysis of the test image demonstrates the capability of the
proposed method to determine the pitch angle of the spiral and to resolve the
difference of 5\degr\ despite the exponential background and small-scale bright
sources.

\section{The Data} 
\label{sec:data}

\begin{figure*}[htbp] 
\centering 
\begin{minipage}[c]{0.47\textwidth} \centering
	\includegraphics[width=\textwidth]{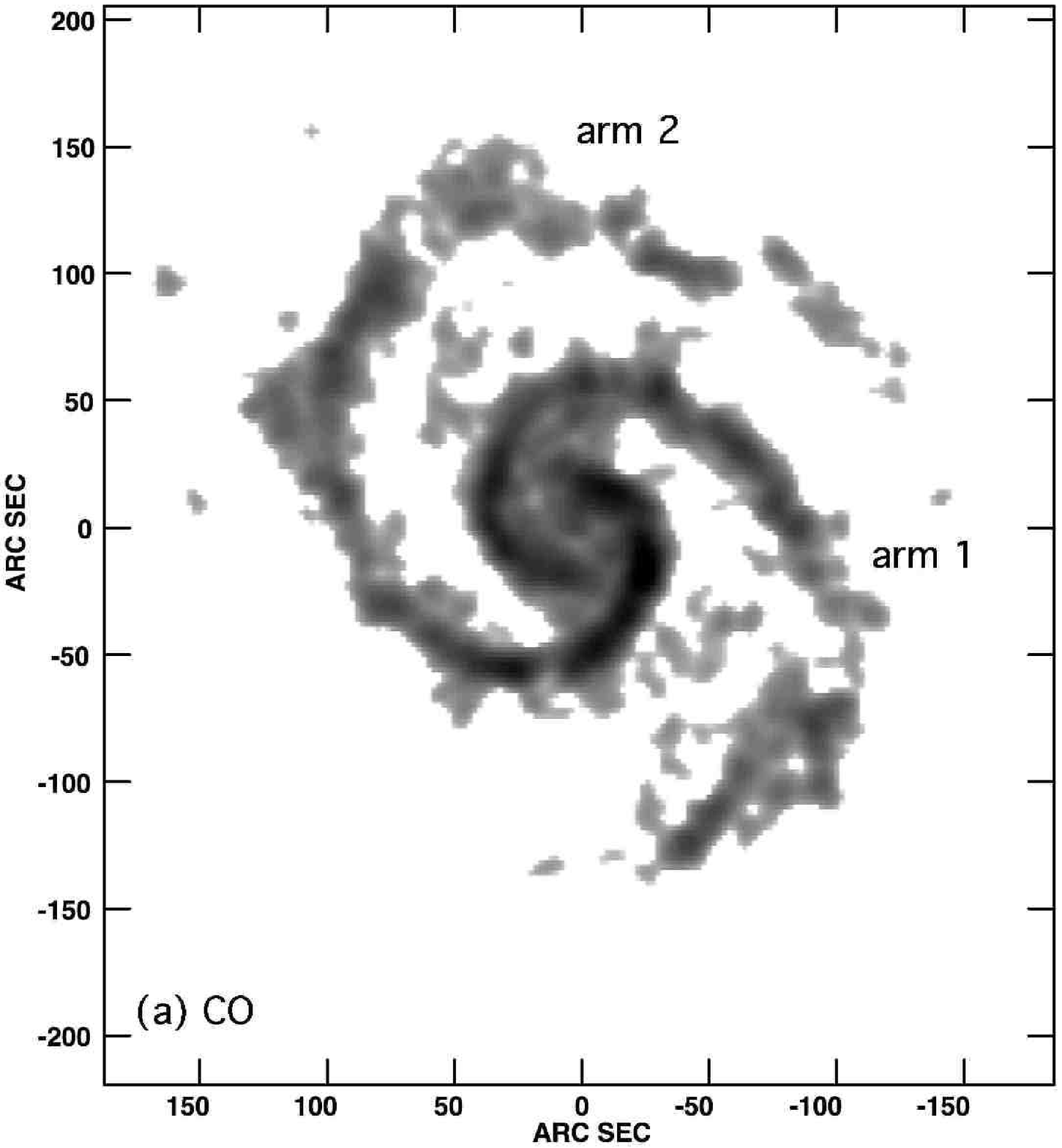} 
\end{minipage}
\begin{minipage}[c]{0.47\textwidth} \centering
	\includegraphics[width=\textwidth]{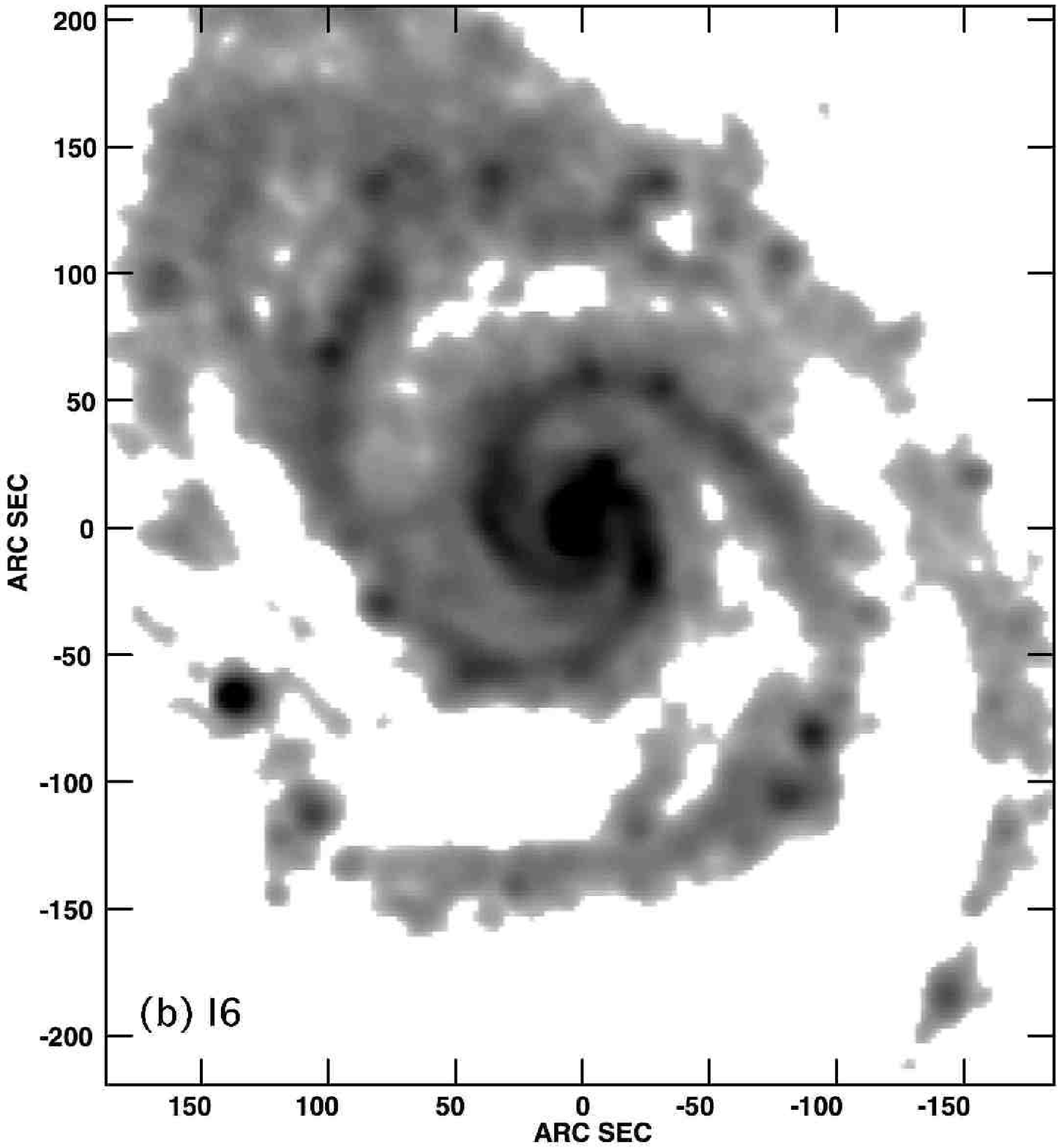} 
\end{minipage}\\[10pt]
\begin{minipage}[c]{0.47\textwidth} \centering
	\includegraphics[width=\textwidth]{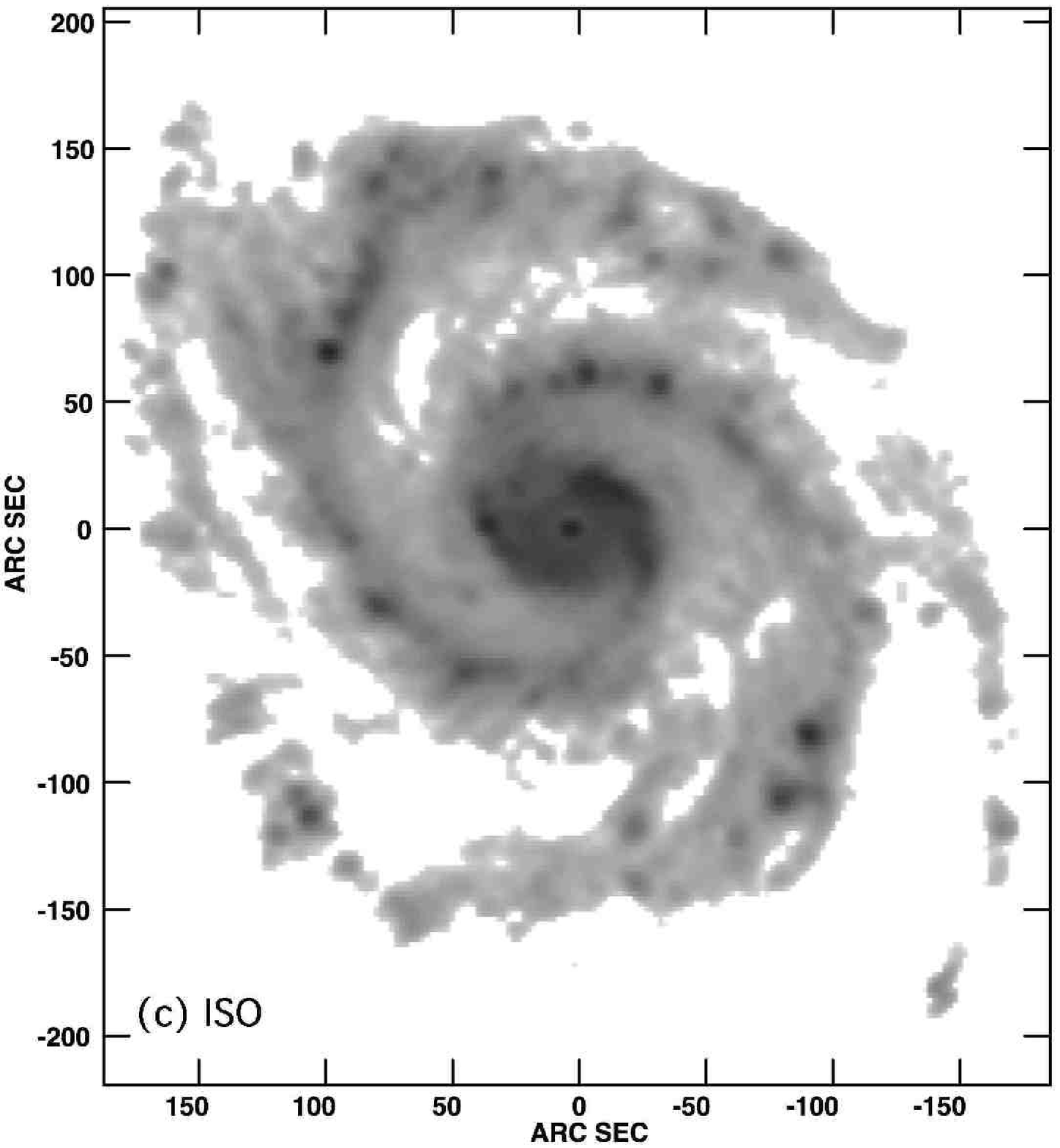} 
\end{minipage}
\begin{minipage}[c]{0.47\textwidth} \centering
	\includegraphics[width=\textwidth]{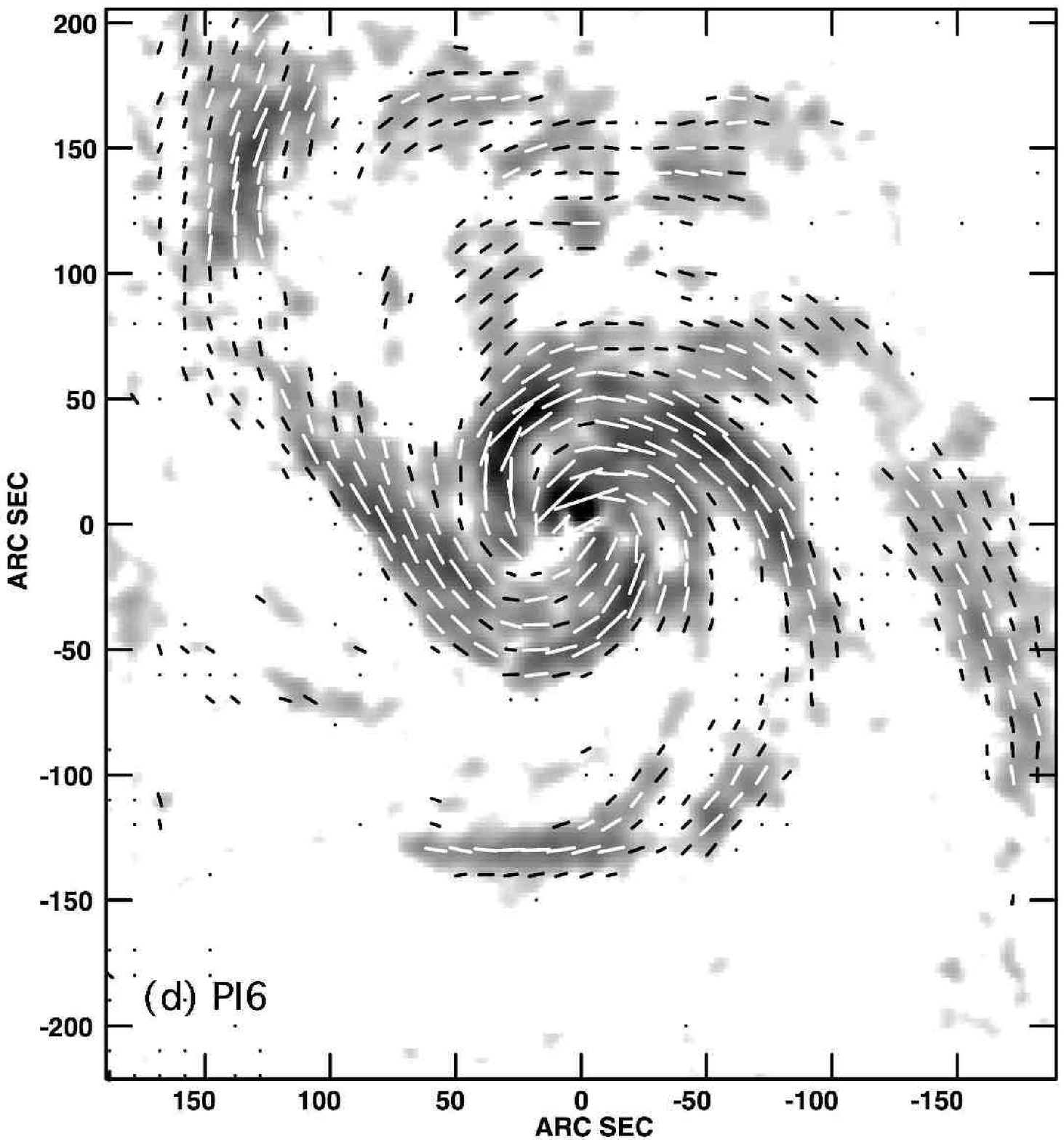} 
\end{minipage} 
\caption{The maps analysed in this paper. \textbf{a)} CO(1--0) emission (Helfer
et al.~\cite{Helfer03}). \textbf{b} Total radio emission at \wav{6} (Fletcher et
al. in preparation). \textbf{c)} Infrared emission at $15\mum$ (Sauvage et
al.~\cite{Sauvage96}). \textbf{d)} Linearly polarized \wav{6} radio emission
(Fletcher et al. in preparation), with B-vectors showing the orientation of
the regular magnetic field (corrected for Faraday rotation). A grey-scale
proportional to the square root of the intensity has been used to emphasise the
spiral structures. Arms 1 and 2 are indicated on the CO map. The resolution is
$8\arcsec$ in all maps and all have been rectified to a face-on orientation.
The co-ordinate system is in seconds of arc relative to the galaxy centre (RA, Dec
(2000):13h 27m 46s, +47$\degr$ 27$'$ 10$''$).}
\label{fig:orig_maps} 
\end{figure*}

M51 is a well observed galaxy and data in many spectral ranges are available.
For our analysis we use four maps: CO($1-0$) line emission of molecular gas,
total and linearly polarized radio continuum emission at \wav{6.2}, and
mid-infrared dust emission at 15$\mu$m. 

The CO($1-0$) line emission at $\lambda2.6$~mm was observed with the BIMA
interferometer and corrected for missing spacings with single dish observations
from the NRAO 12m dish on Kitt Peak (Helfer et al.\ \cite{Helfer03}). CO
emission is the best available tracer of the molecular gas in M51.

It is often assumed that dust lanes are the best tracers of large scale spiral
arm shocks, although recent Hubble Space Telescope images show a plethora of
dust-absorption spurs and branches attached to the M51 spiral arms that clearly
require a more subtle interpretation. The recent $850\mum$ map of M51 at
15\arcsec\ resolution --- twice the resolution of the data used in this paper
--- by Meijerink et al.~(\cite{Meijerink05}) shows that most of the (cold) dust
in M51 originates from an exponential disc, rather than from narrow spiral arms
associated with the dust-absorption lanes seen in the optical bands. We note
that at high resolution the strongest CO($1-0$) emission appears to closely follow
the main spiral dust lanes in the inner $r\lesssim 4\kpc$ (Aalto et
al.~\cite{Aalto99}).

The \wav{6.2} total and linearly polarized radio continuum emission was observed
with the VLA \footnote{The VLA is operated by the NRAO. The NRAO is a facility
of the National Science Foundation operated under co-operative agreement by
Associated Universities, Inc.} and corrected for missing short baselines using
maps obtained with the Effelsberg \footnote{The Effelsberg telescope is operated
by the Max-Planck-Institut f\"ur Radioastronomie on behalf of the
Max-Planck-Gesellschaft.} telescope (Fletcher et al. in preparation). The
total emission is a mixture of synchrotron radiation of cosmic-ray electrons
spiralling in interstellar magnetic fields and bremsstrahlung emission from
thermal electrons. The relative fractions of synchrotron and thermal emission
vary with position; at the location of $\HII$ regions in the spiral arms the
emission is up to $~50\%$ thermal but in the inter-arm regions this fraction
falls to $\le20\%$. Hence, the total radio intensity I6 is mostly a measure of
the total strength of the magnetic field and the cosmic ray electron density.
The polarized emission PI6 is purely non-thermal, and its intensity is a measure
of the strength of the regularly oriented magnetic field.\footnote{A note on
terminology is required. In this paper we use the term "regular magnetic field"
to refer to the magnetic field that gives rise to the observed polarized
synchrotron emission. Polarized emission can be produced by an
\emph{anisotropic} random magnetic field as well as a coherent mean field; in
this work we ignore the distinction except for a brief discussion in
Sect.~\ref{sec:discuss}.} The polarization B-vectors were corrected for Faraday
rotation using new \wav{3.5} data (Fletcher et al. in preparation) and show
the orientation of the regular magnetic field. The similarity of the polarized
intensity maps at \wwav{3.5}{6.2} (apart from generally reduced intensities at
\wav{3.5} in line with the non-thermal spectral index) and the low Faraday
rotation measures mean that Faraday depolarization does not significantly effect
the PI6 map.

The $15\mum$ dust emission (denoted ISO throughout this paper) was observed with
ISOCAM (Sauvage et al.\ \cite{Sauvage96}) and is a combination of thermal
continuum emission from dust particles and line emission from PAH molecules.

All maps were smoothed to 8\arcsec\ resolution, equivalent to $\simeq400\pc$ at
the assumed distance of M51 (9.7~Mpc, Sandage \& Tammann \cite{Sandage74}). The
maps were rectified to a face-on orientation using an assumed inclination of
$20\degr$ (where $0\degr$ is face-on) and a position angle of $-10\degr$ for the
orientation of the major axis (Tully \cite{Tully74}). Figure~\ref{fig:orig_maps}
shows the face-on maps at 8\arcsec\ resolution.

If the de-projection to face-on orientation uses incorrect parameters and the
spiral arms have a simple mathematical description, such as logarithmic spirals,
sine-wave like oscillations will appear in log(radius)--azimuth plots of
the ridges (see Fig.~2 of Kennicutt~\cite{Kennicutt81}). For $r<6\kpc$ we do not
find any such oscillations in either arm, but for $r>6\kpc$ a series of 3 peaks
and troughs is present in the I6 and ISO of arm~1 (Fig.~\ref{fig:log_ridges}).
The pattern may indicate a distortion of the disc in this region of M51, rather
than intrinsic variability in the spiral pattern. Without detailed information
on the velocity field we cannot investigate this possibility; however, whatever
the cause of the oscillations, the \emph{relative} positions of the I6 and ISO
arms will be the same.

\section{Results} 
\label{sec:results} 

\subsection{The anisotropic wavelet transform} 
\label{sec:anisotropic}

\begin{figure*}[htbp] 
\centering 
\begin{minipage}[c]{0.47\textwidth} \centering
	\includegraphics[width=\textwidth]{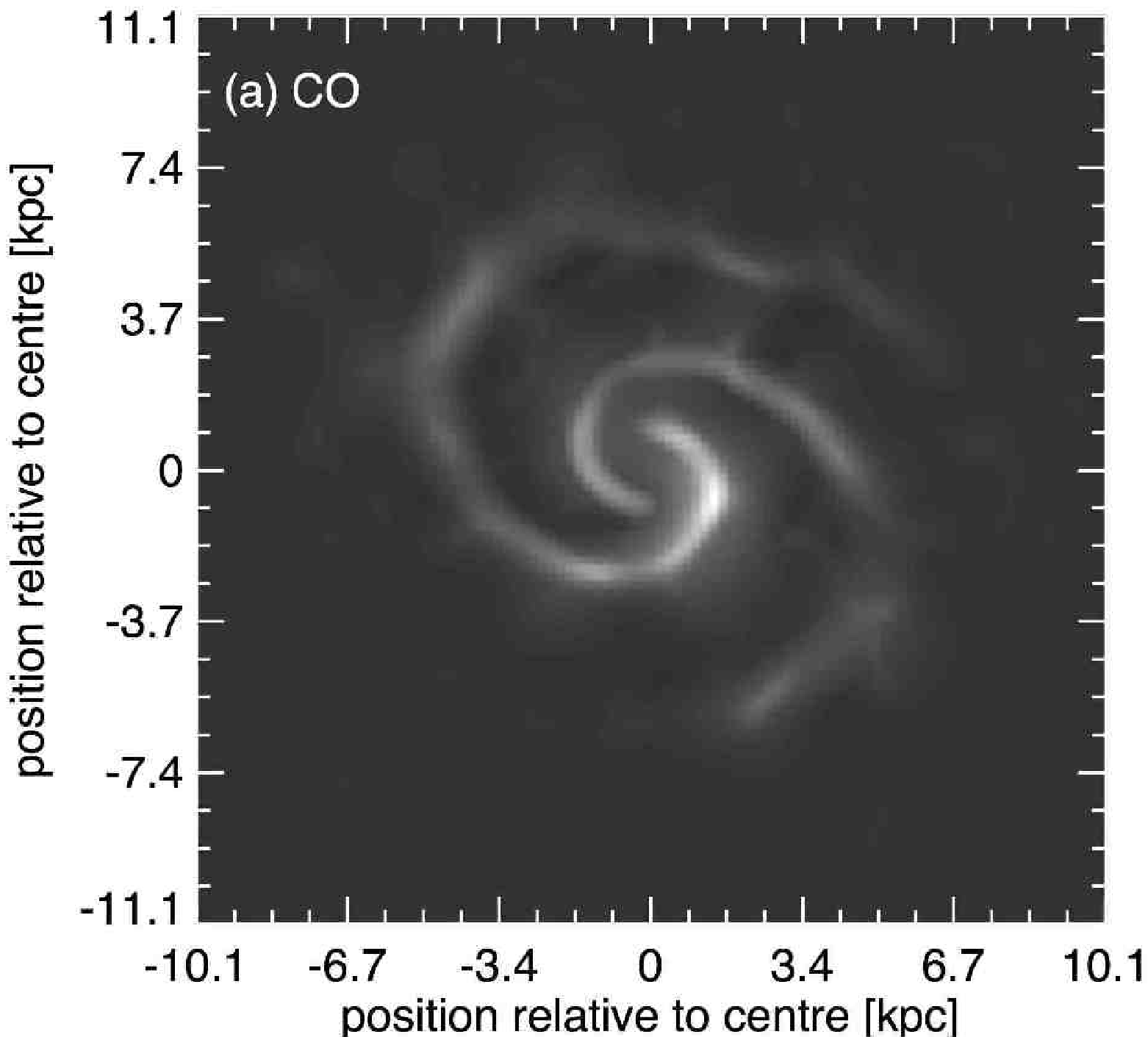} 
\end{minipage}
\begin{minipage}[c]{0.47\textwidth} \centering
	\includegraphics[width=\textwidth]{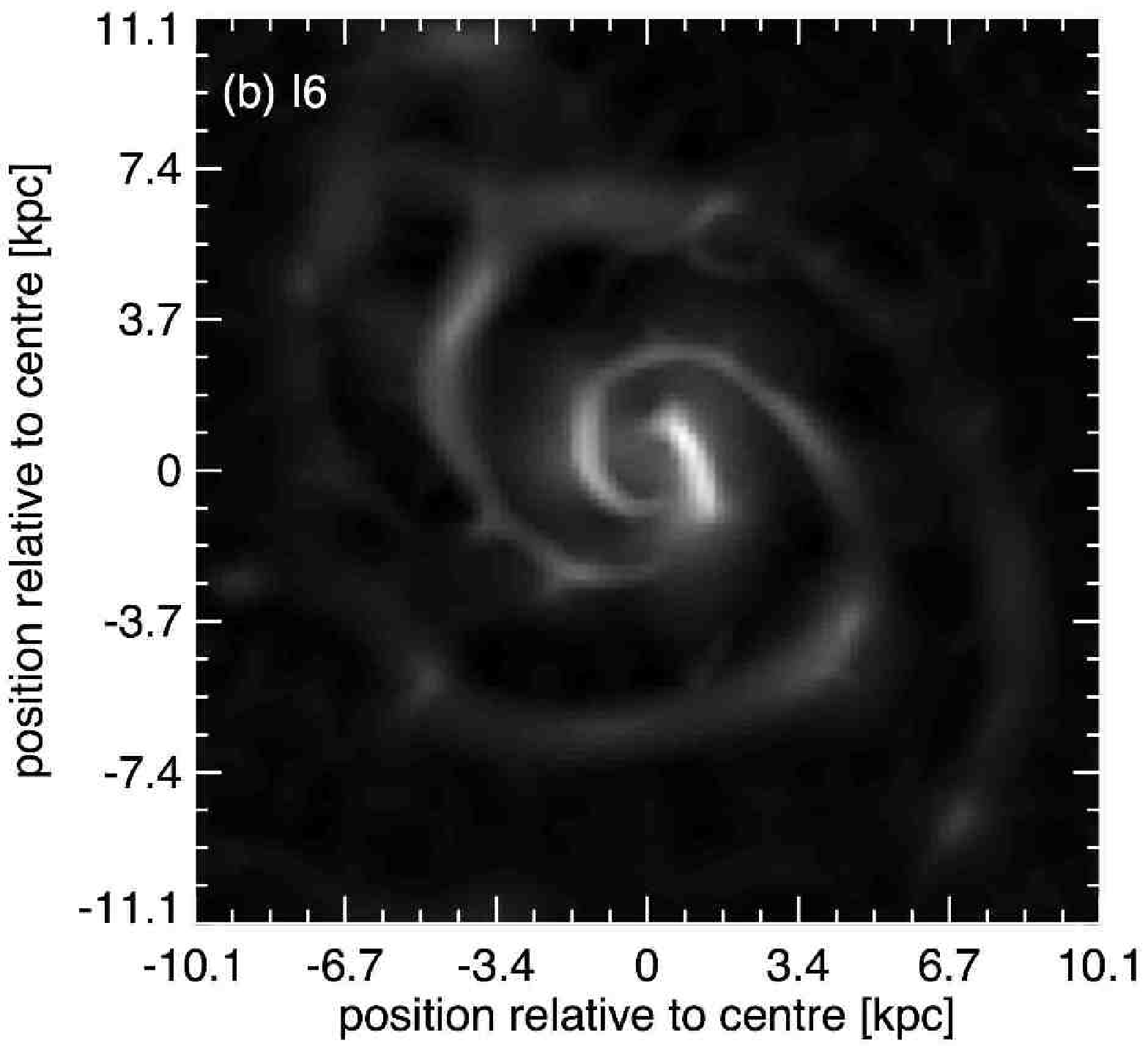} 
\end{minipage}\\[10pt]
\begin{minipage}[c]{0.47\textwidth} \centering
	\includegraphics[width=\textwidth]{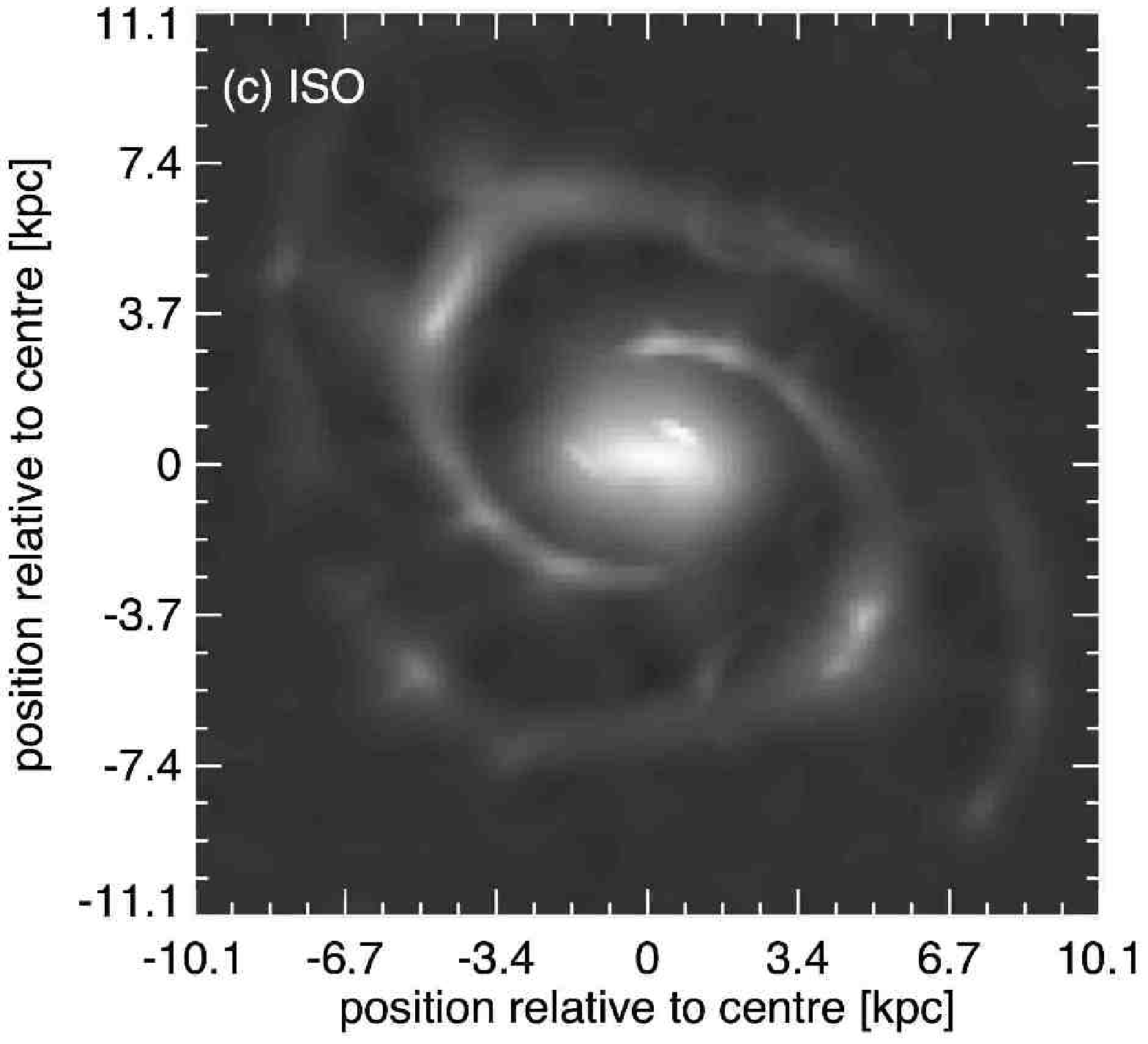} 
\end{minipage}
\begin{minipage}[c]{0.47\textwidth} \centering
	\includegraphics[width=\textwidth]{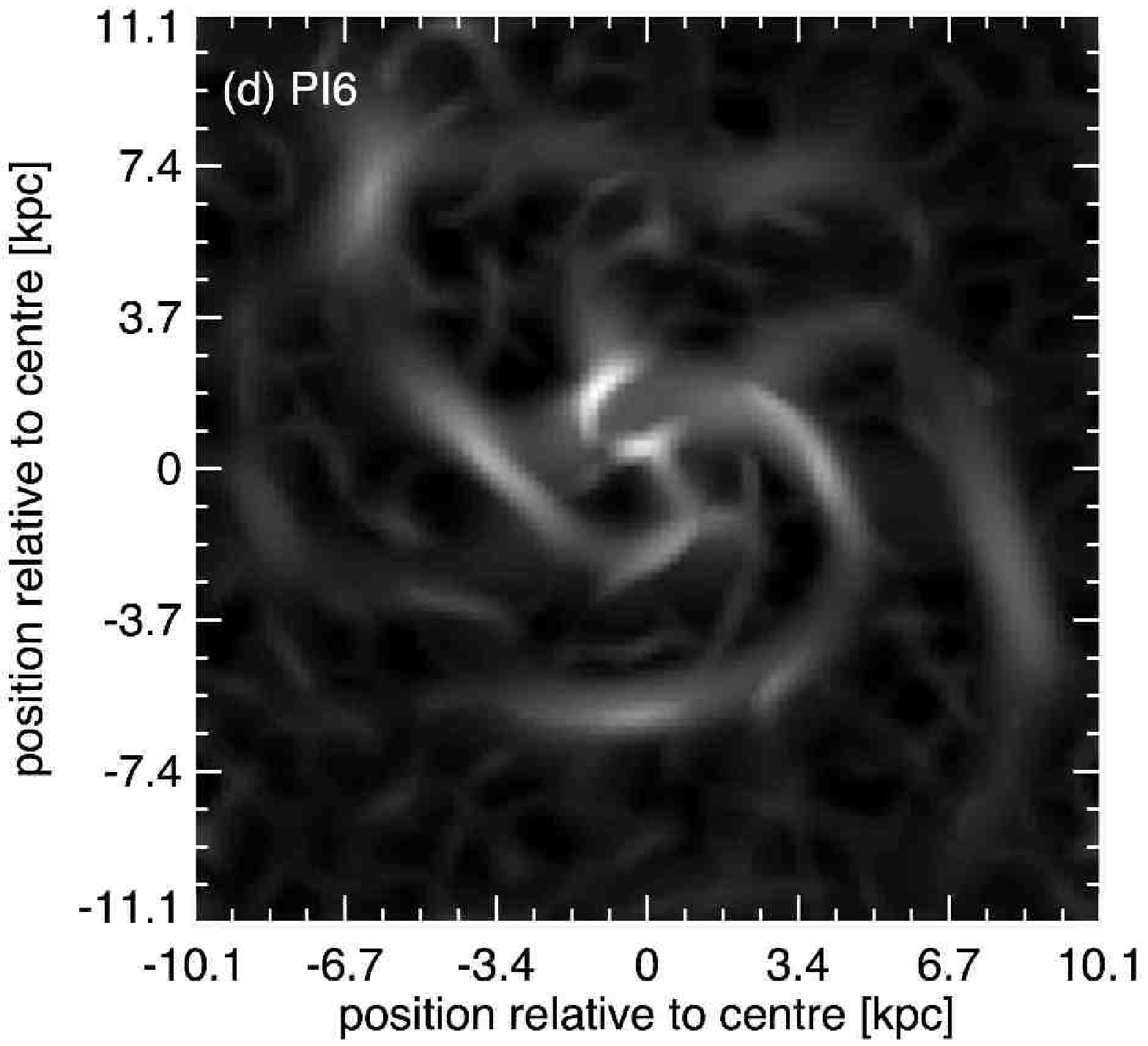} 
\end{minipage} 
\caption{The maximum anisotropic-wavelet coefficients of the original maps shown
in Fig.~\ref{fig:orig_maps}, corrected for inclination of the galaxy.
\textbf{a)} CO(1--0) emission. \textbf{b)} total radio emission at \wav{6}.
\textbf{c)} infrared ISO $15\mum$ emission. \textbf{d)} polarized \wav{6} radio
emission. The grey-scale is proportional to the amplitude of the wavelet
coefficient (white is high). At each location the wavelet coefficient was
calculated for a fixed range of scales and $180$ orientations, the highest
coefficient is shown so the wavelet scale may be different between pixels.} 
\label{fig:wav_coeff} 
\end{figure*}

Figure~\ref{fig:wav_coeff} shows the maximum wavelet coefficients for the four
maps in Fig.~\ref{fig:orig_maps}. The scale parameter $a$ of the wavelet was
fixed to the approximate width of the spiral arms; tests showed that the results
are insensitive to the aspect ratio of the wavelet in the range $0.5$--$2$ times
the chosen aspect ratio. The anisotropic wavelet transform clearly picks out the
elongated arms in the images. In addition, smaller scale structures such as
spurs roughly perpendicular to the arms are identified. In the PI6 map the low
signal to noise ratio (relative to the other maps) and the patchiness of the
diffuse emission results in a web of structures showing up in
Fig.~\ref{fig:wav_coeff}(d), however extended arms of polarized emission are
also evident.

The location of the wavelet coefficient maxima at a given radius are interpreted
as the position of the spiral arm ridge. To ensure an even spacing of measured
positions we select an initial point that clearly lies on a spiral arm and then
choose the location with the maximum wavelet coefficient that is a fixed
increment in azimuth away from the initial position (2 degrees) within a arc of
fixed opening angle. The anisotropic wavelet will pick out spurs and
ridges that are connected to the spiral arms (Fig.~\ref{fig:wav_coeff}) but by
selecting the maximum wavelet coefficient within an arc we attempt to pick out
the continuous structure of the spiral arm. The pitch angle of the arm is given
by the orientation of the anisotropic wavelet that generates the maximum wavelet
coefficient at each of these positions.

Uncertainties in the measured pitch angle are caused by (i) the form of the real
structures (they can be curved, asymmetric, having varying width etc.) and (ii)
by instrumental noise in the map. In our approach structures on scales smaller
than the scale of the arms -- such as discrete bright clouds, spurs and breaks
-- can be considered as noise. To calculate the errors in our measured pitch
angles we used a Monte-Carlo technique. We define all structures on scales less
than the approximate width of the spiral arms of $1\kpc$ as belonging to the
noise. We carry out large-scale wavelet filtration of the image, keeping only
scales $a < 1\kpc$ and consider this to be a map of the noise. Then we reanalyse
the original map overlaid with the noise map shifted (the noise map can be
laterally shifted, rotated or reflected) by an arbitrary amount and calculate
the pitch angles again. The process is repeated using several arbitrarily
shifted noise maps to obtain the standard deviation for the pitch angle of the
arms.

\subsection{The location of the spiral arms} 
\label{sec:armpos}

\begin{figure*}[htbp] 
\centering
\begin{minipage}[b]{12cm}
	\centering
	\includegraphics[width=\textwidth]{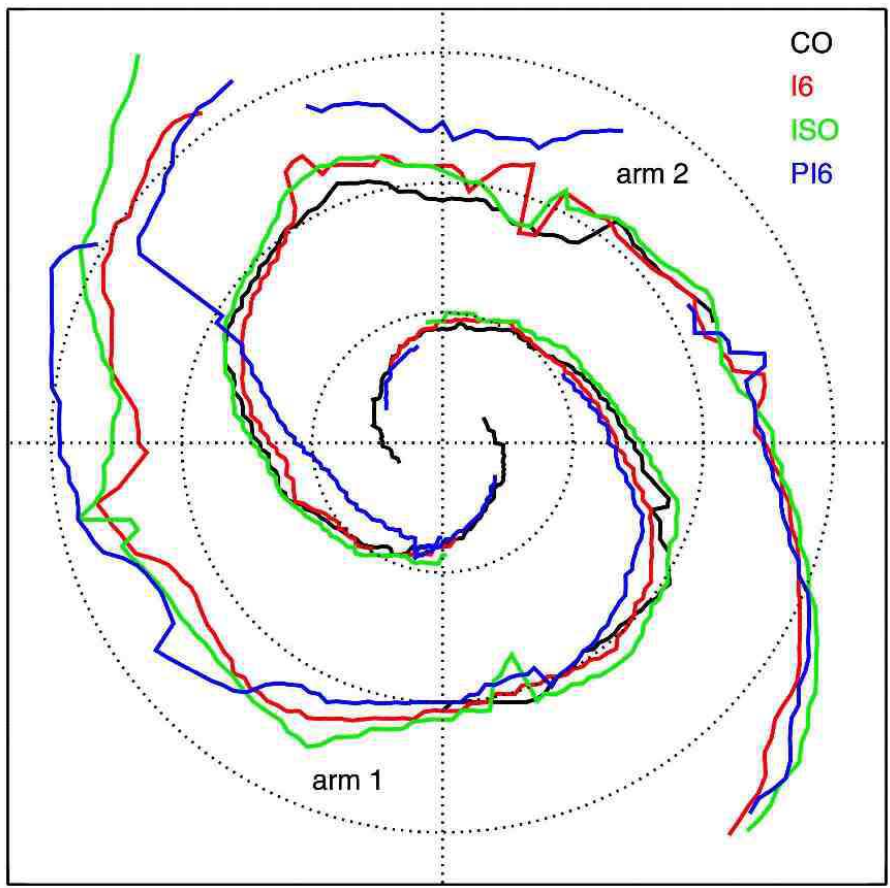}
	\par\vspace{0pt}
\end{minipage}
\hfill
\begin{minipage}[b]{5.5cm}
\centering
	\caption{Location of the various spiral arm ridges in the plane of M51; dotted
	lines show the galacto-centric radii 3, 6 and 9$\kpc$. I6 and PI6 are the total
	and polarized radio intensities at \wav{6}, ISO the $15\mum$ infrared emission.
	The ridges are the positions where the anisotropic wavelet coefficient has a
	maximum, tracking along the spiral arms, as described in the text.} 
	\label{fig:ridges}
	\par\vspace{0pt}
\end{minipage}
\end{figure*}

\begin{figure*}[htbp] 
\centering
\includegraphics[width=0.45\textwidth]{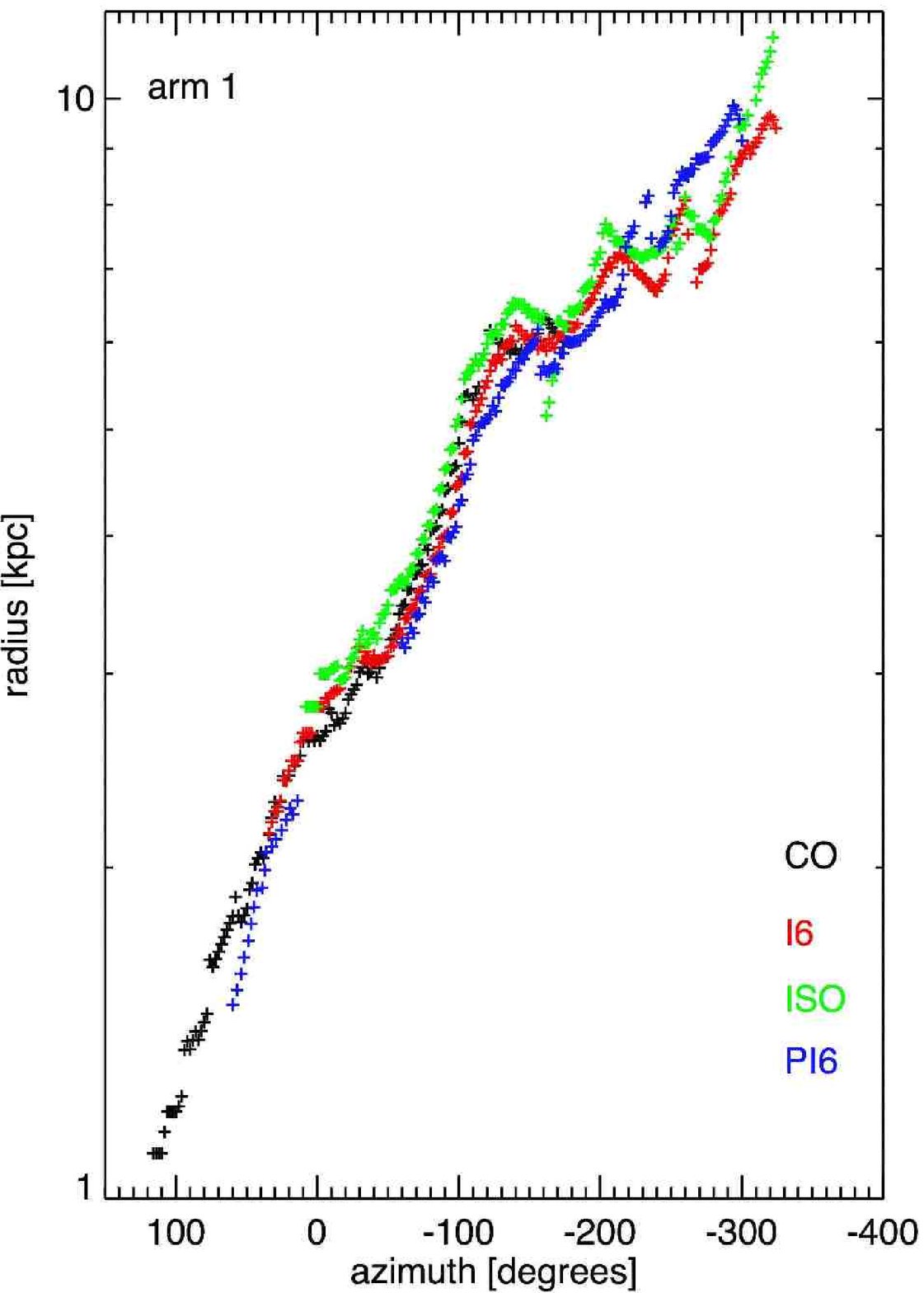}
\includegraphics[width=0.45\textwidth]{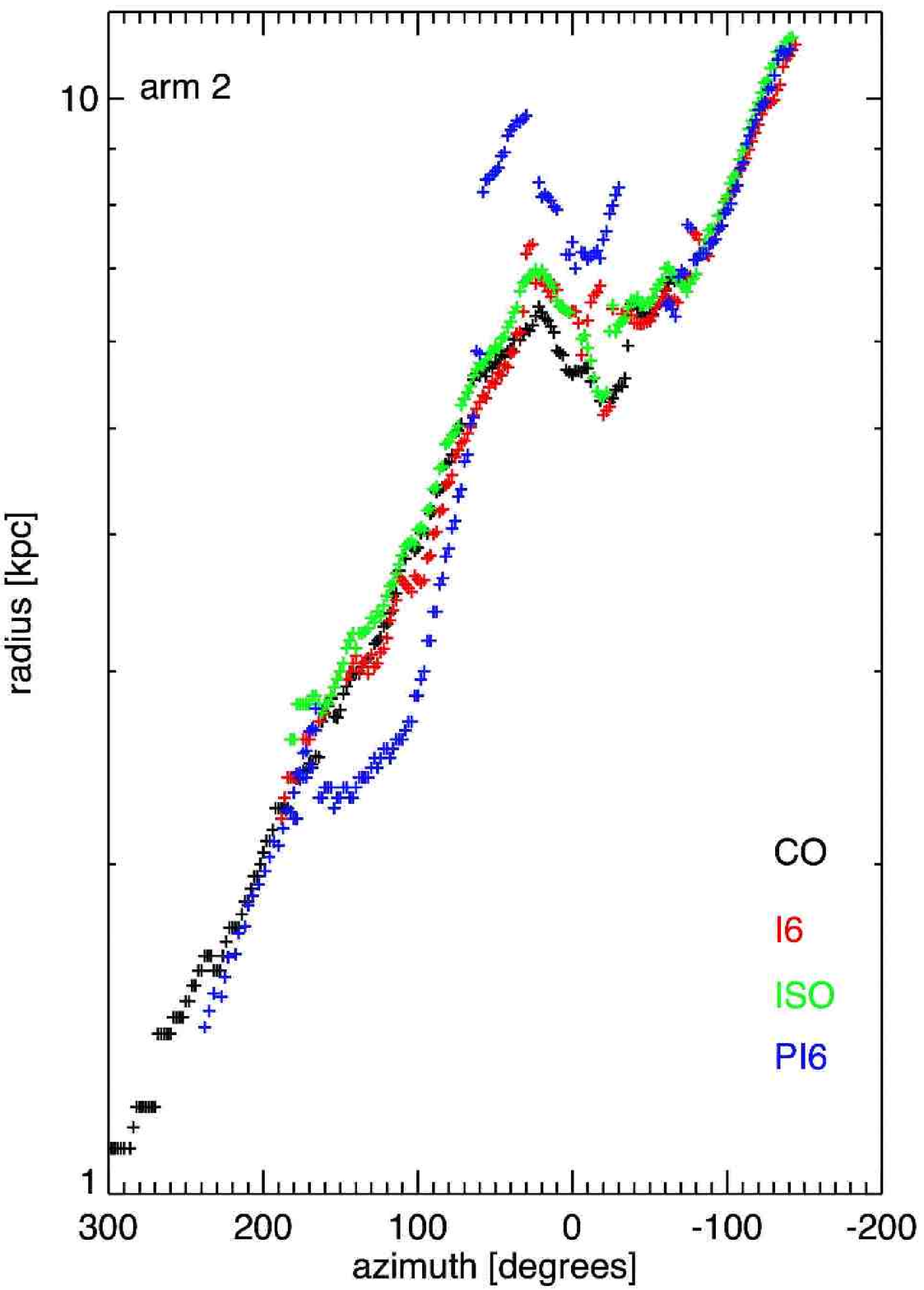} 
\caption{Radius (on a logarithmic scale) versus azimuth plots of the spiral arm
ridges. This shows the regions of systematic offsets in ridges quite well
(compare to Fig.~\ref{fig:ridges}). The azimuthal angle is measured
counter-clockwise from the top of the face-on galaxy.} 
\label{fig:log_ridges} 
\end{figure*}

Figure~\ref{fig:ridges} shows the location of the spiral arm ridges in M51
derived from the anisotropic wavelet transform (Fig.~\ref{fig:wav_coeff}) of the
total and polarized \wav{6}, CO and ISO $15\mum$ maps. An alternative view of
the same data --- in log(r)--azimuth co-ordinates --- is shown in
Fig.~\ref{fig:log_ridges}. 

The small-amplitude 'wiggles' in the plotted ridge lines are
due to the fixed number of positions where the ridge position was measured and
do not translate into pitch angle fluctuations. 

There are three systematic offsets between the spiral arms traced in different
wavebands; the most obvious offset is that of the polarized emission, but since
this is the most complicated to discuss we will leave it to last.

First, the CO spiral arms are consistently situated on the concave side of the
ISO $15\mum$ arms. The shift between the two ridges is typically $\sim 100\pc$,
increasing to around $600\pc$ in the northern part of arm 2. This offset fits
neatly with the chronology of the conversion of gas into stars in the
large-scale shock model of spiral arms (Roberts~\cite{Roberts69}, Shu et
al.~\cite{Shu72}, Tosa~\cite{Tosa73}). The CO emission traces molecular gas
clouds, whose formation is triggered by a shock on the upstream side of the
spiral arm (the concave side within the co-rotation radius where the gas rotates
faster than the spiral pattern). After dense cores in these clouds have
collapsed, stars form downstream of the position where the clouds formed. The
ISO $15\mum$ emission comprises line emission from PAH molecules and a continuum
component from warm dust particles heated by Ly-$\alpha$ and UV radiation, the
prime source of which are young star clusters that will have moved further
downstream of the spiral arm shock. Figures~\ref{fig:ridges} \&
\ref{fig:log_ridges} clearly show, for the first time that we are aware of, the
systematic offset of dense gas spiral arms and infrared arms along regions several
kpc long.

Second, the I6 spiral arms are generally offset on the concave side of
the ISO arms, by $200$-$300\pc$. There is also a tendency for I6 to sit
upstream of the CO arms inside $\sim 6\kpc$ radius, by around $100$-$200\pc$.
This offset is harder to explain. Thermal radio emission will be enhanced by the
same new star clusters that heat the dust and PAHs. Hence no offset between the
I6 and ISO arms is expected from thermal effects. The most significant
non-thermal process (on $\sim 100\pc$ scales) associated with star formation is
probably an increase in total magnetic field strength, caused by turbulent
tangling of magnetic field lines. At our 8\arcsec\ resolution ($\sim 400\pc$)
this synchrotron emission will be predominantly unpolarized, again tending to
align the I6 and ISO ridges. The shift of the I6 arms upstream of the ISO
arms can be due to the polarized synchrotron emission.

The polarized ridge in the inner part of arm~2 is clearly located in the
inter-arm region and may have no direct connection with the gaseous arms (see
below), but along most of the inner part of arm~1, $r\lesssim 6\kpc$, the
polarized ridge is close to the CO ridge, sometimes very closely aligned but
often shifted about $300\pc$ upstream. This displacement of the PI6 arm~1
indicates that the regular magnetic field is strongest upstream of the CO arm,
and the PI6 ridge is shifted further upstream than the I6 ridge. At radii of
6--8~kpc the PI6 ridges of arm~1 and arm~2 move from the concave to the convex
side of the other tracers. The shift was also seen in the \wwav{18}{20}
polarization maps of Horellou et al.~(\cite{Horellou92}). Beyond this radius,
the two polarization arms behave differently: along the outer part of arm~2 the
polarization ridge is intertwined with the I6 and ISO, whereas along the outer
part of arm~1 the PI6 sits consistently on the convex side of the I6 and ISO
ridges. CO is absent at these large radii. The differences between the inner and
outer spirals are difficult to explain, but may be related to the conjunction of
an inner spiral density wave and an outer material spiral pattern identified by
Elmegreen et al. (\cite{Elmegreen89}), overlapping in the range $6\kpc\lesssim
r\lesssim 8\kpc$. The largest shifts between PI6 and other tracers occur on the
side of M51 closest to the companion galaxy, which may have caused the two outer
spirals (Howard \& Byrd \cite{Howard90}).

The offset of the PI6 ridge upstream of the CO ridge in arm~1 in the inner disc
indicates that the spiral shock lies slightly upstream of the CO ridges. The
position of the I6 ridge is then due to a balance between the polarized
(non-thermal) emission pulling it upstream of the CO and the thermal and
unpolarized emission pulling it towards the ISO ridge.

Third, radio polarization arms are partly situated in inter-arm regions, most
noticeably over $\sim 3\kpc$ in the inner part of arm~2 and north of the middle
part of arm~2 and in the outer part of arm~1, where they are displaced by around
$700\pc$ from the other arms. Polarized intensity maps at \wwav{3}{6} (Fletcher
et al. in preparation) clearly show concentrations of emission in these
inter-arm regions, particularly in the inner part of arm~2: we emphasise that
these elongated polarized structures are \emph{not} artefacts caused by the
wavelet analysis. These sections of the PI6 ridges resemble the magnetic arms
observed in NGC~6946 (Beck \& Hoernes~\cite{Beck96}). However, the PI6 ridge
that is associated with arm~2 near the galaxy's centre forms a \emph{continuous
structure} that is initially well aligned with the CO ridge then, as the radius
increases, moves into the inter-arm region, rejoins arm~2 and then crosses
inter-arm space again to link up with arm~1. This interlacing of large scale
magnetic spiral structures and gaseous arms is quite unlike the behaviour of the
magnetic arms seen in NGC~6946, but resembles that in NGC~2997 (Han et
al.~\cite{Han99}). Higher resolution and sensitivity may reveal similar
structures in other galaxies. Its explanation will require more sophisticated
modelling of the interplay between spiral density waves and magnetic fields than
has been used to date and is beyond the scope of this paper.

\subsection{Variation of pitch angles along the spiral arms}
\label{sec:armpitch}

\begin{figure}[htpb] 
\centering
\includegraphics[width=0.45\textwidth]{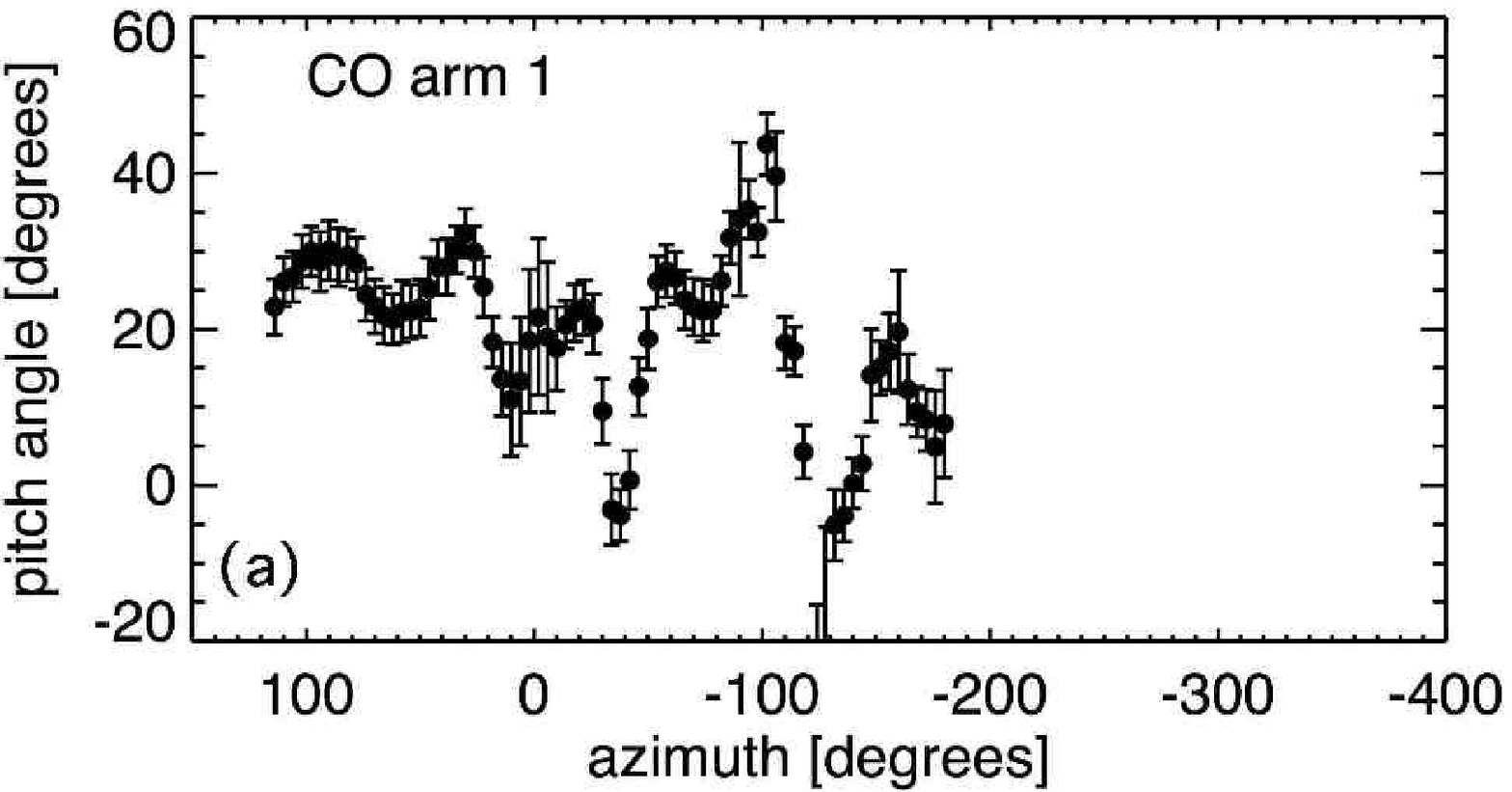}
\includegraphics[width=0.45\textwidth]{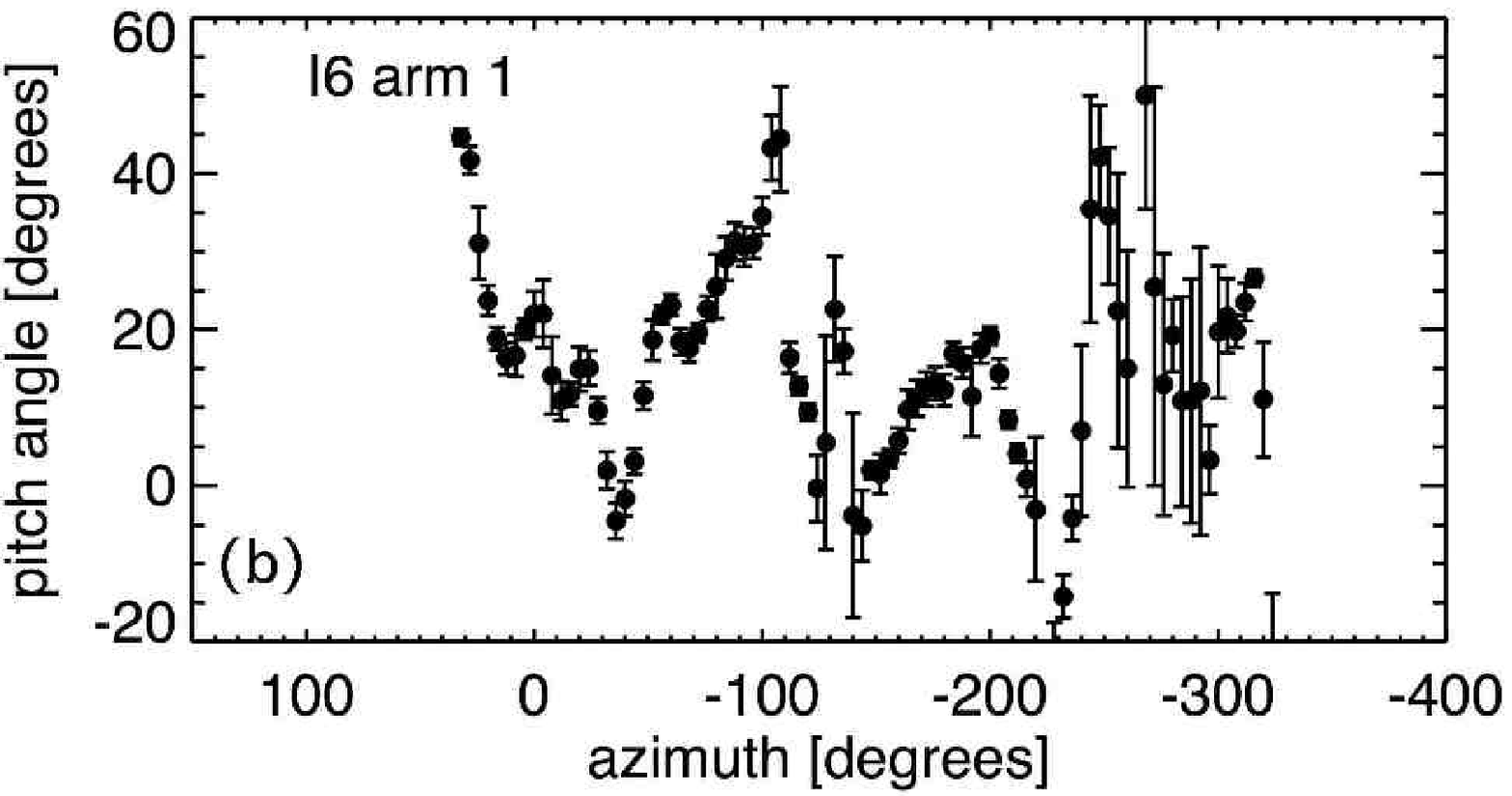}
\includegraphics[width=0.45\textwidth]{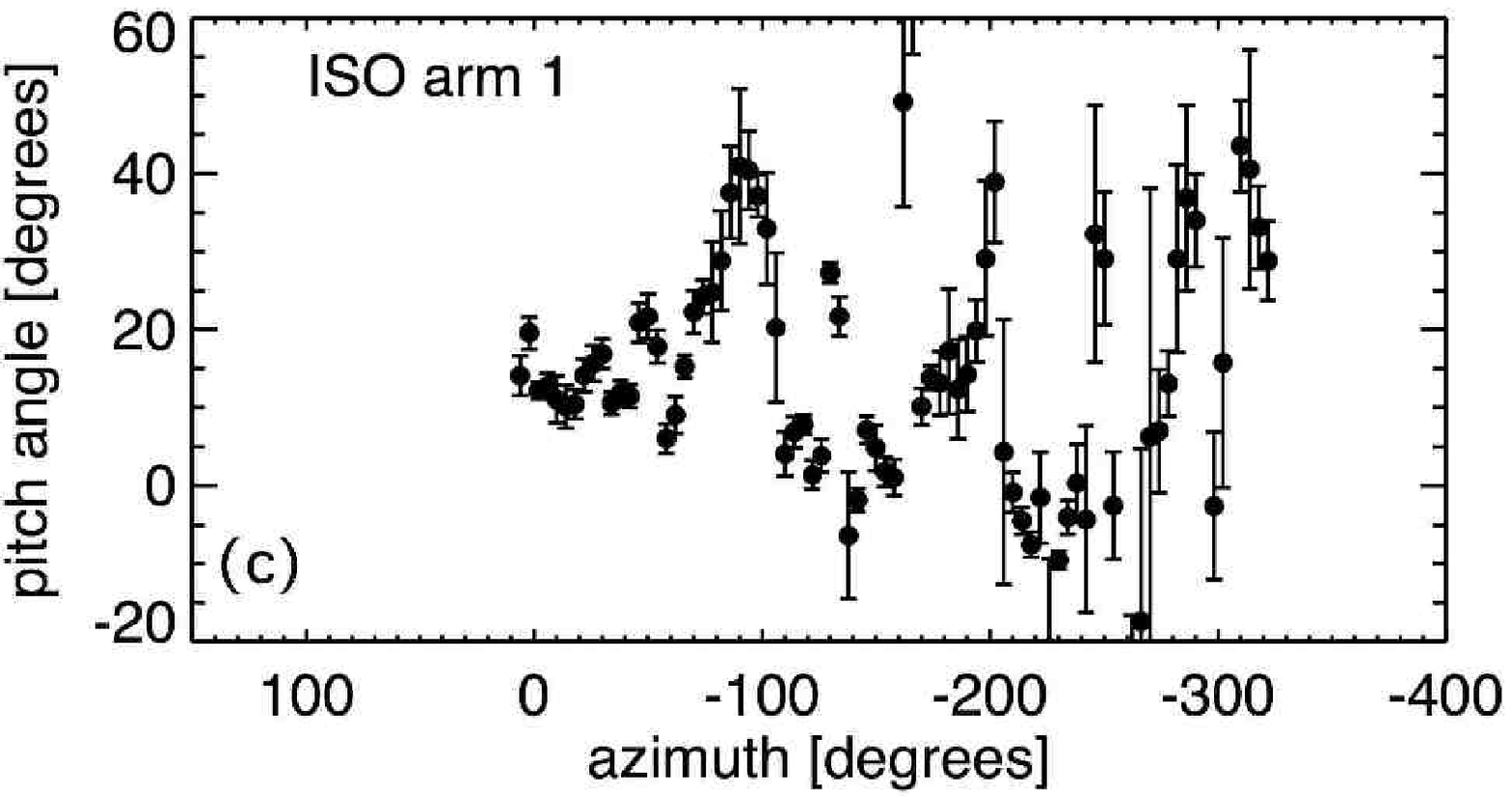}
\includegraphics[width=0.45\textwidth]{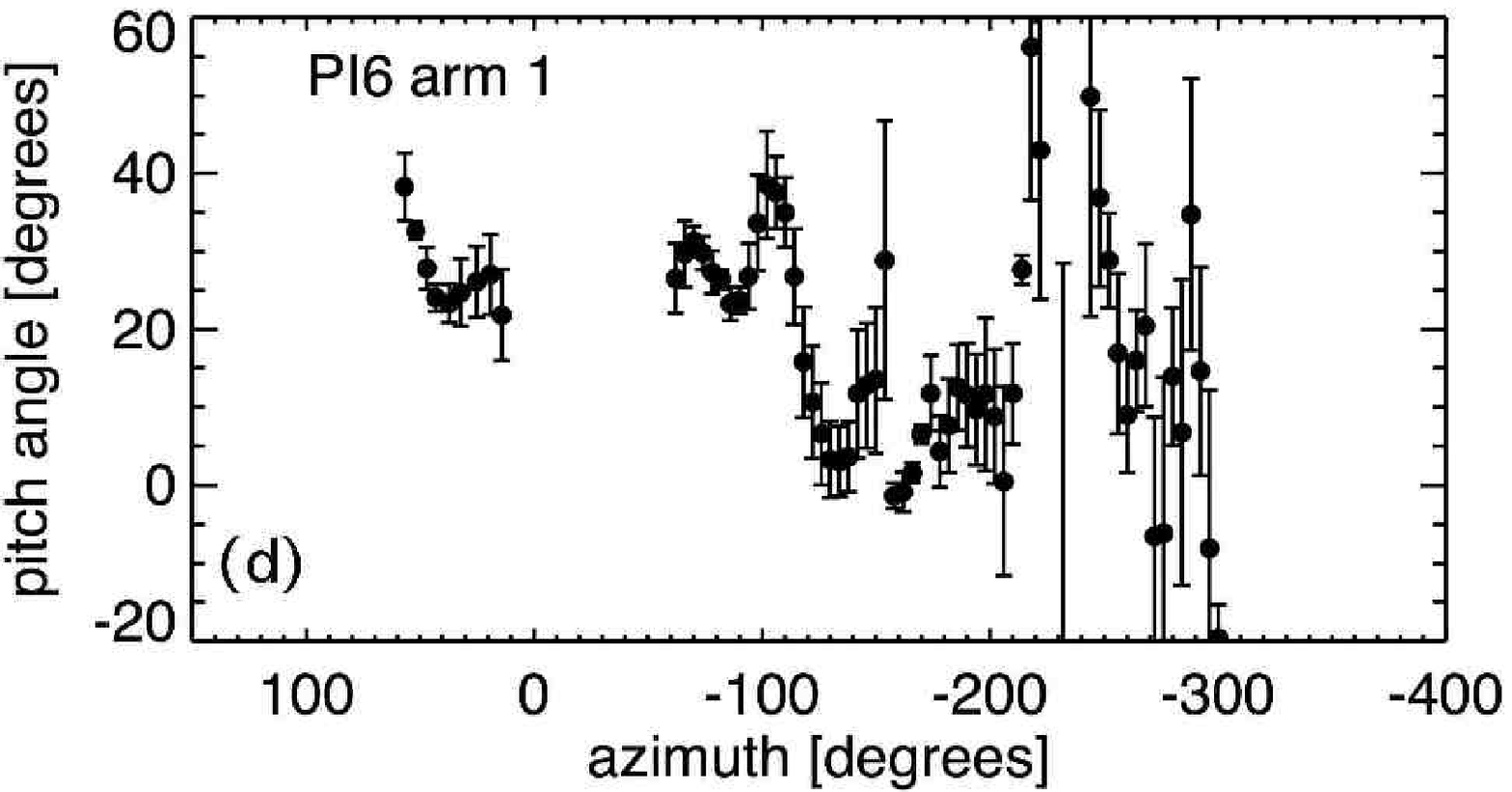} 
\caption{Pitch angles of arm~1 (see Fig.~\ref{fig:orig_maps} for arm labelling).
\textbf{a)} CO(1--0) emission. \textbf{b)} I6 \wav{6} total emission.
\textbf{c)} ISO $15\mum$ emission. \textbf{d)} PI6 \wav{6} polarized emission.
Every $6^{\circ}$ in azimuth (measured counter-clockwise from the top of the
face-on galaxy) along an arm, the pitch angle of the wavelet with the highest
amplitude is shown. Errors are calculated according to the Monte Carlo method
described in Section~\ref{sec:anisotropic}. The arms run from left (smallest radius)
to right (largest radius) in all panels.}
\label{fig:pitch_arm1} 
\end{figure}

\begin{figure}[htpb] 
\centering
\includegraphics[width=0.45\textwidth]{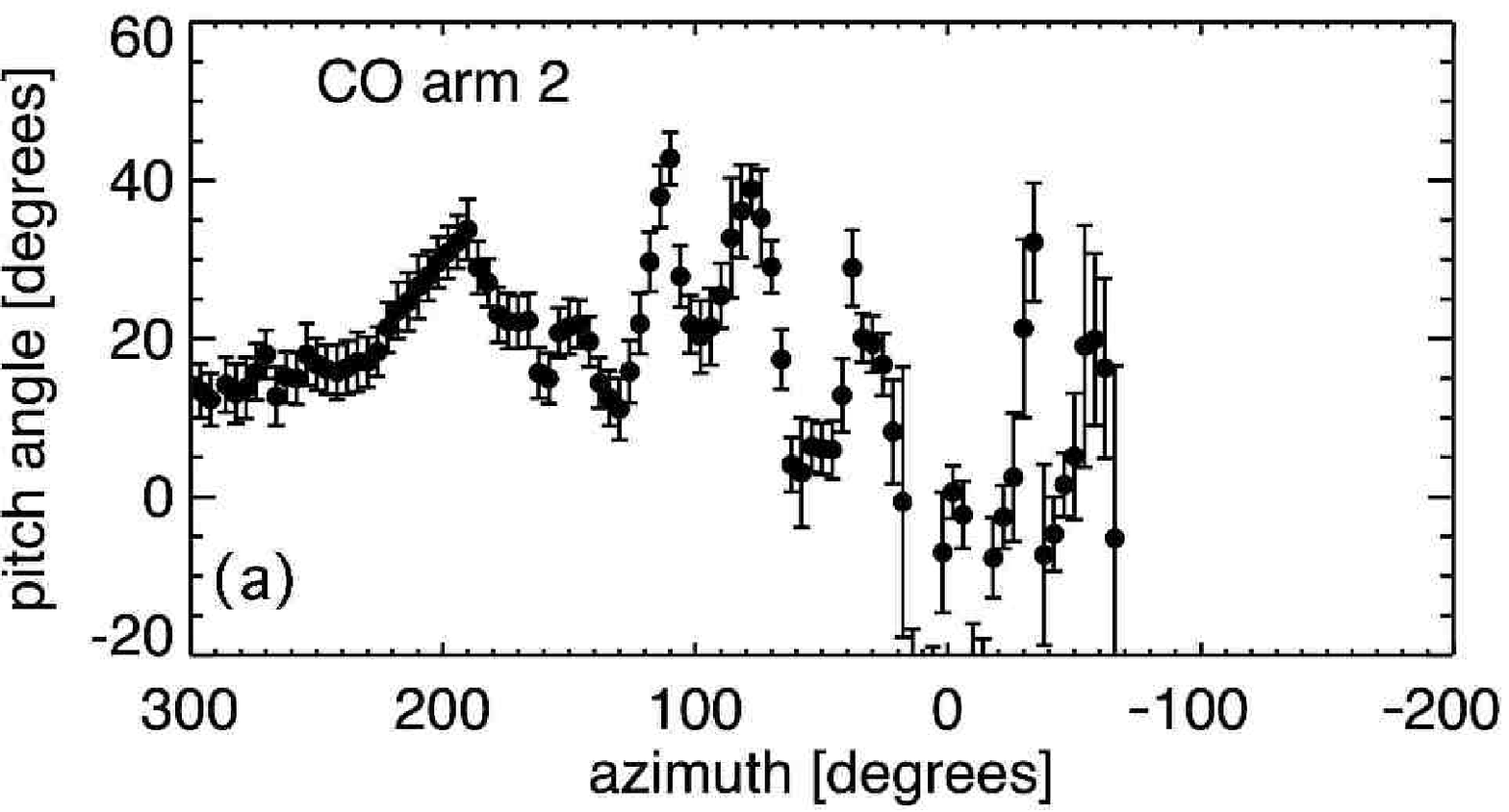}
\includegraphics[width=0.45\textwidth]{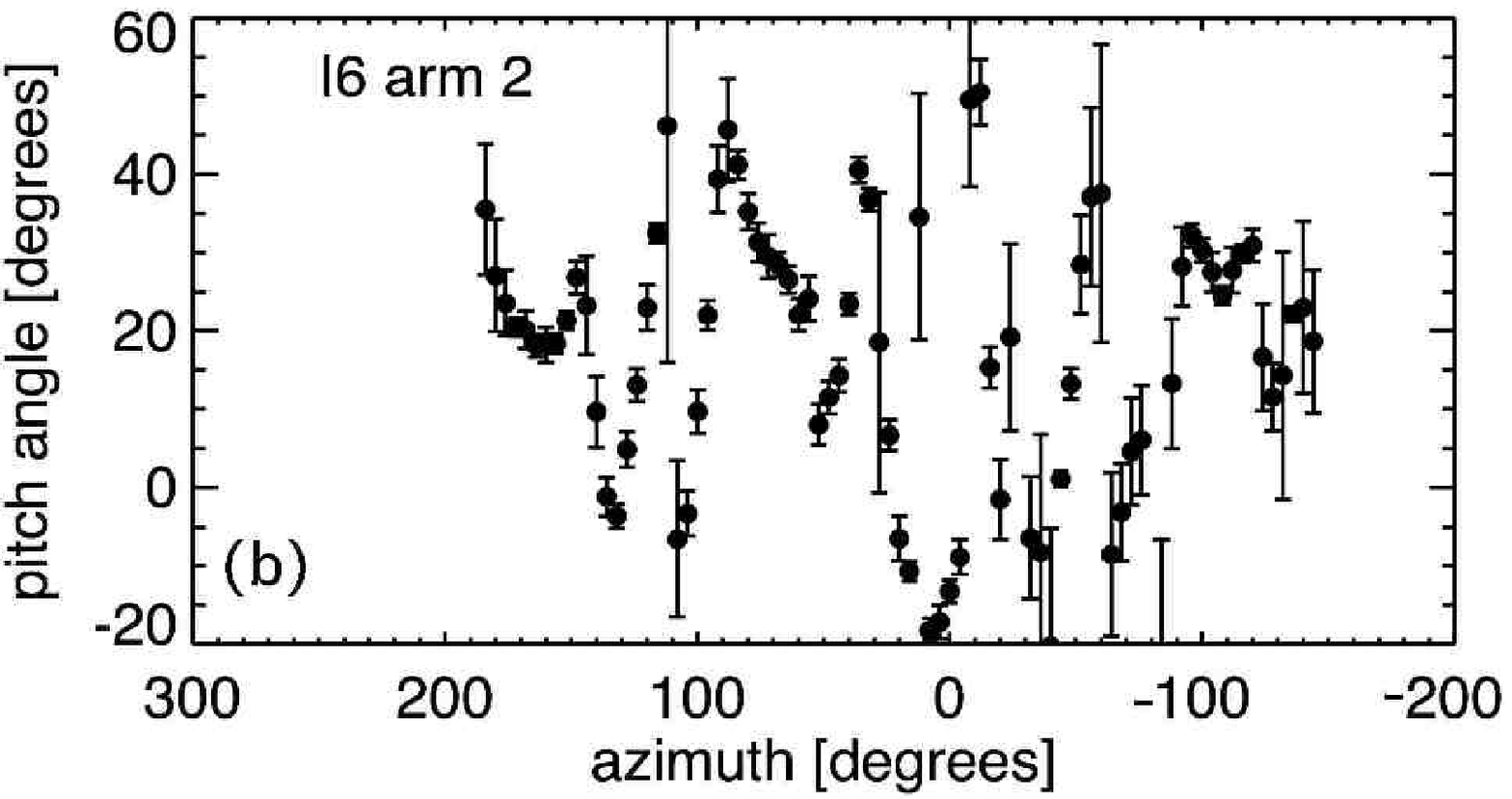}
\includegraphics[width=0.45\textwidth]{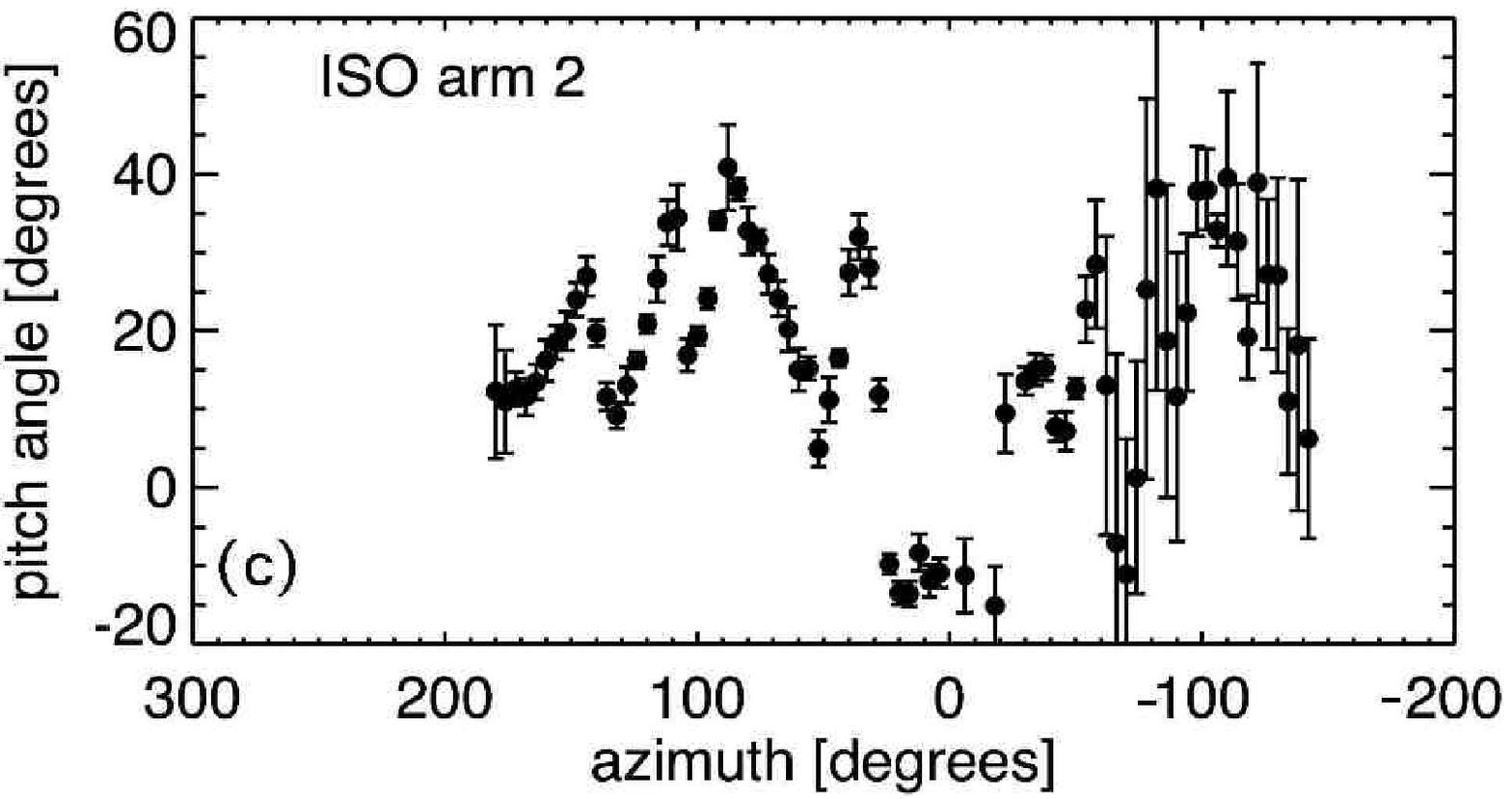}
\includegraphics[width=0.45\textwidth]{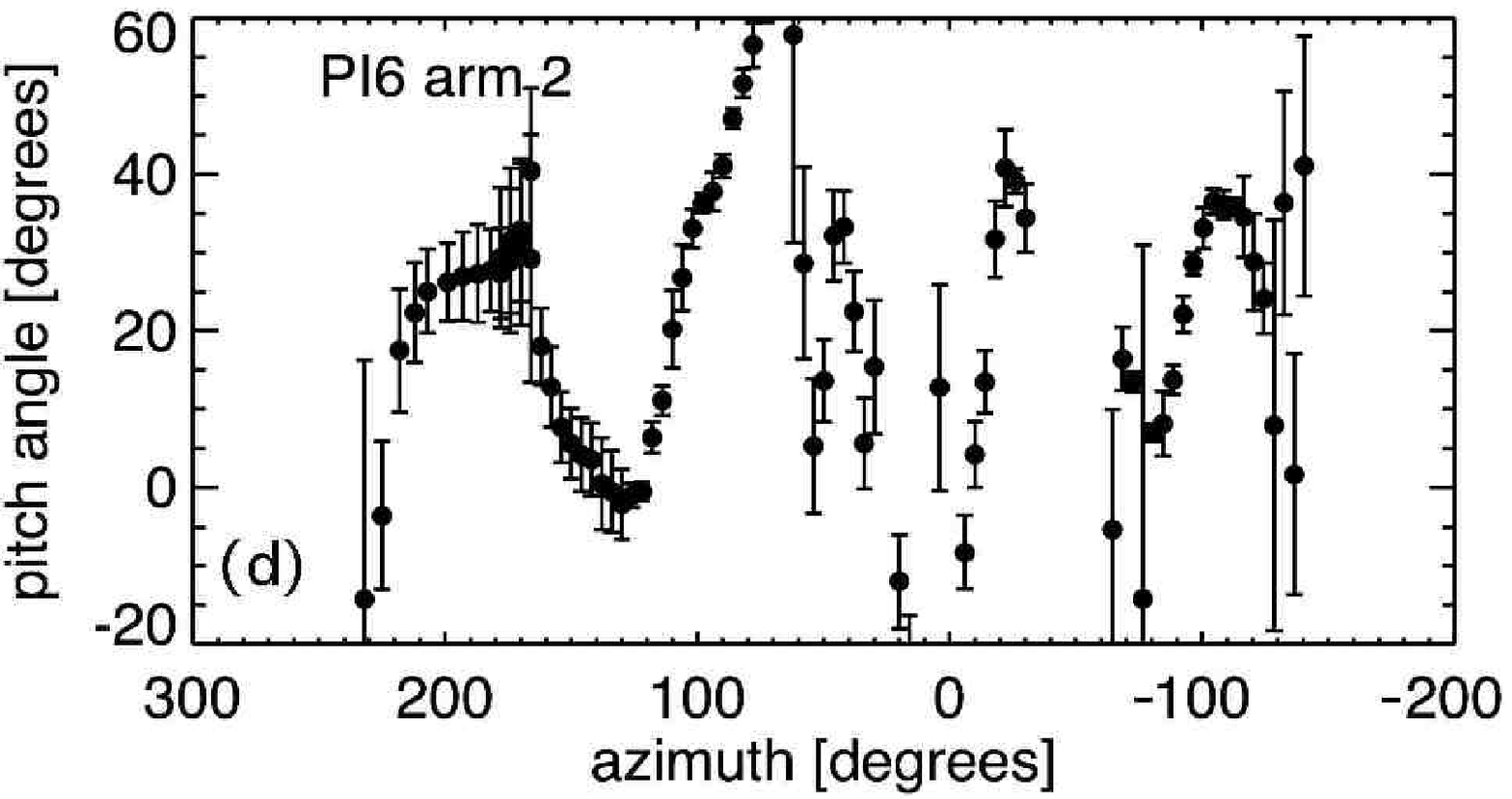} 
\caption{As in Fig.~\ref{fig:pitch_arm1} but for arm~2.} 
\label{fig:pitch_arm2} 
\end{figure}

Figures~\ref{fig:pitch_arm1} and \ref{fig:pitch_arm2} show how the pitch angles
of the spiral arms vary along their length. The uncertainties in pitch angles
tend to get greater with increasing distance along the arms; the generally
fainter emission (compared to the inner disc) means that the maximum wavelet
coefficients are smaller and the arms become less well defined
(Fig.~\ref{fig:wav_coeff}). First we will discuss the pitch angles for each
spiral arm tracer in turn, then we will compare the pattern of pitch angles in
different tracers.

The two CO spiral arms can be traced through $\Delta\phi\approx 360\degr$ in
azimuth. In both CO arms the pitch angle in the inner quarter turn
($\Delta\phi\approx 90\degr$) is approximately constant, within the errors, but
the pitch angles of the two arms in this region differ by $\sim 10\degr$: in
arm~1, $\pco\approx 25^{\circ}$ and in arm~2, $\pco\approx 15\degr$. Further
along the arms the pitch angles follow different trends. Thus, despite the
apparent similarity of the two CO arms, they are not symmetric under rotation.
There is substantial variation of $\pco$ in both arms, with ranges
$0\degr\lesssim\pco\lesssim 30\degr$ in arm~1 and $-20\degr\lesssim\pco\lesssim
40\degr$ in arm~2. There is a tendency for low or negative pitch angles to occur
toward the ends of both arms. This shift in pitch angles to low and even
negative values occurs at $\phi\approx -100\degr$ in arm~1 and $\phi\approx
+60\degr$ in arm~2. The transition occurs around the place where the character
of the CO arms -- especially arm~2 -- markedly changes in the original map
(Fig.~\ref{fig:orig_maps}). The galacto-centric radii of these positions are
around $6\kpc$, which is the approximate position of the co-rotation radius
(Elmegreen et al.~{\cite{Elmegreen92}). At these azimuths the pitch angles 
of the other tracers show similar variations. Furthermore in the azimuthal ranges
$40\degr>\phi>-100$ in arm~1 and $200\degr>\phi>60\degr$ in arm~2 the maxima and 
minima in $\pco$ follow each other with the same interval in $\phi$; changes in 
$\pco$ occur symmetrically along the two arms, but with different magnitudes.  

The \wav{6} total radio emission (I6) and ISO $15\mum$ spirals are both quite
well defined over the first half turn ($\Delta\phi\approx 200\degr$) with
systematic trends in their pitch angles. The variation in pitch angles is
stronger than in the CO arms and the uncertainty in the outer arm pitch angles
is higher; this is because the arms are better defined in CO, whereas in I6 and
ISO the arms are broader and contain bright patches, probably arising from
regions of intense star formation. The \wav{6} polarization arms (PI6) are
generally the least well defined of all, although the first $\Delta\phi\approx
130\degr$ of arm~2 is very clear and shows a sharp and systematic variation in
pitch angle from $\ppi \simeq 30\degr$ to $\ppi \simeq 0\degr$ and back up to
$\ppi \simeq 60\degr$. Both PI6 arms show strong variations in pitch angle -- a
range of $-20\degr\lesssim\ppi\lesssim 60\degr$ in arm~2.

In Section~\ref{sec:armpos} we saw that the position of the spiral arm ridges is
not the same in all tracers, but that there can be small but systematic shifts
in the arm location. Now we consider how the pitch angles compare along arms.

In general there is quite good agreement and $\pco\approx\pii$, particularly in
arm~1 for $\phi>-120\degr$ where the pitch angles agree within errors for
$\Delta\phi\approx 100\degr$. In arm~2 there are alternating regions of good and
bad agreement in pitch angles. Overall, the similarity in pitch
angles and locations where both tracers show changes to higher or lower pitches,
the physical process producing the CO arms is likely the same as that producing
the I6 arms, as expected for density wave compression.

Finally we compare $\pii$ and $\piso$. In arm~2 the angles are for the most part
equal, and where they are not the tendency (i.e. whether there is a decrease or
an increase in pitch angle) is the same. Thus, despite the systematic small
offset in the location of the I6 and ISO arms, their pitch angles are apparently
modified for the same reasons.

\subsection{Orientation of the regular magnetic field and pitch angles of the
gaseous and magnetic arms} 
\label{sec:Bpitch}

In this Section we will compare the orientation of the regular magnetic field
\emph{lines} with the pitch angles of the gaseous spiral \emph{structure}
(traced by CO emission) and with the pitch angles of the spiral structure in
polarization (PI6).

The intrinsic orientation of the regular magnetic field ($\pB$) was determined
from the observed Stokes Q and U values by using observations at \wav{3.5} and
\wav{6.2} to correct for Faraday rotation. The errors in magnetic field pitch
angles were calculated using the noise in the observed maps only and do not
include any systematic effects, such as uncertainty in the galaxy inclination
(such systematic effects will equally effect the wavelet derived spiral arm
pitch angles and so should not invalidate our comparisons). We only plot $\pB$
where the polarized intensity at both wavelengths is greater than three times
the noise level. In order to increase the signal to noise ratio and thus to see
the azimuthal trends in $\pB$ better, we have used Q and U data smoothed to
15\arcsec\ (the anisotropic wavelet effectively smoothes the original 8\arcsec\
data in seeking the optimum position and orientation of the spiral arms).
However, we have also calculated $\pB$ at 8\arcsec\ resolution and the main
results --- concerning the alignment or misalignment of $\pB$ with $\pco$ and
$\ppi$ --- do not change.

\begin{figure*}[htbp]
\includegraphics[width=0.45\textwidth]{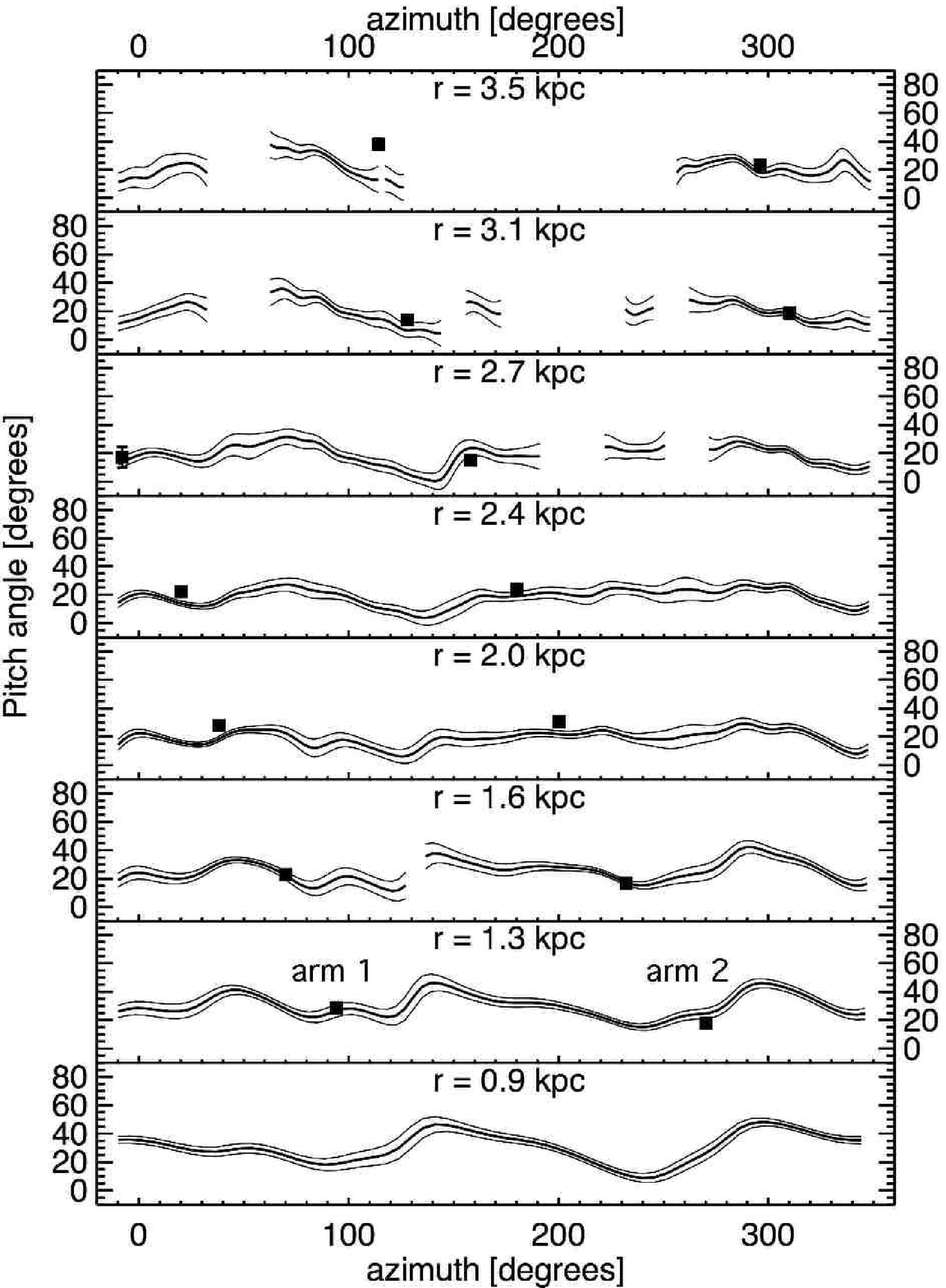} 
\hfill
\includegraphics[width=0.45\textwidth]{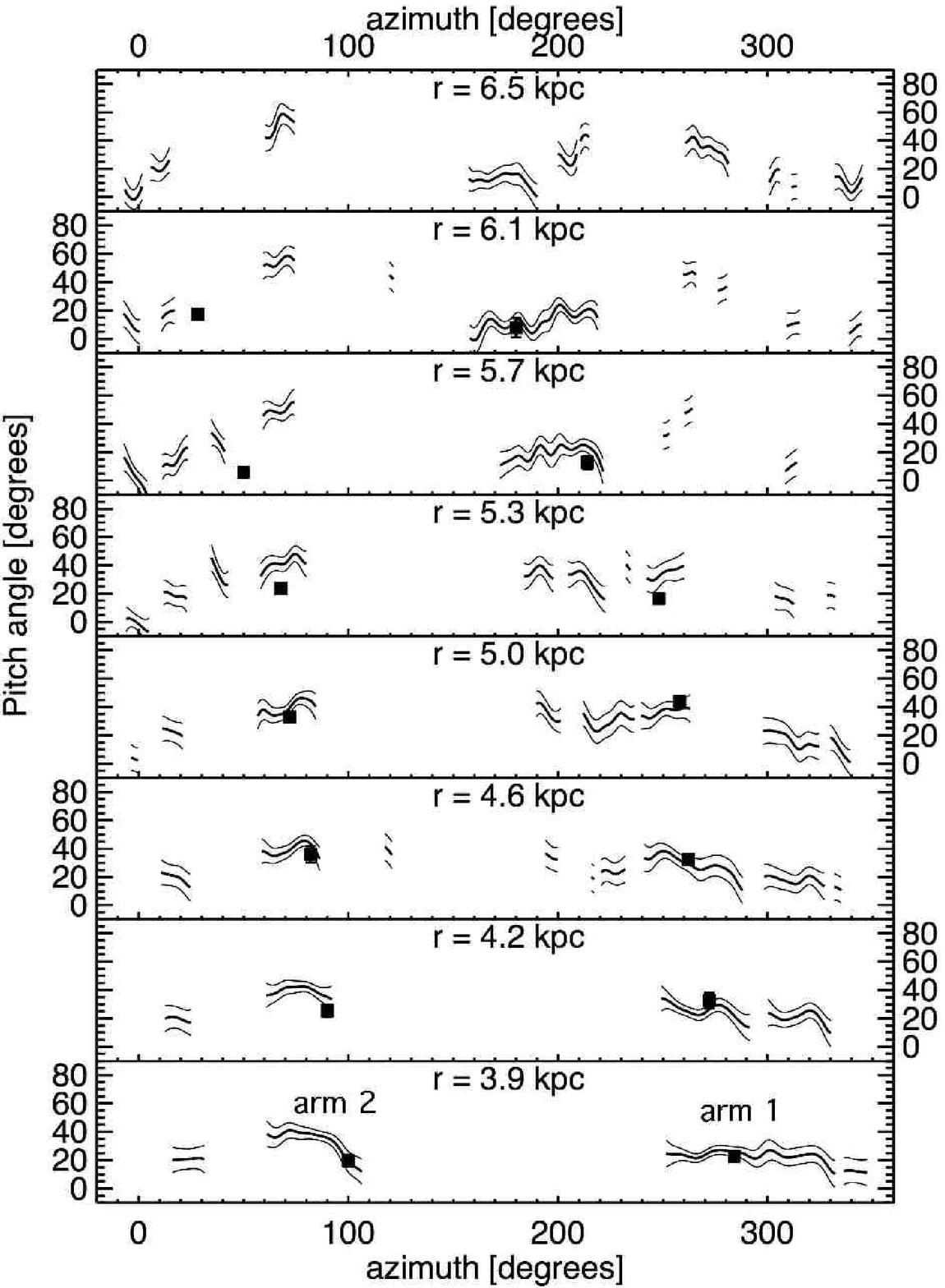} 
\caption{Azimuthal variation of the orientation of the regular magnetic field
$\pB=1/2\arctan(Q/U)$, corrected for Faraday rotation, plotted as curves at
different radii from the centre of M51. Uncertainties are shown by upper and
lower curves. The resolution of the observations is 15\arcsec\. Angles are only
plotted where the polarized intensity at both \wav{3} and \wav{6} is above
3$\times$ the noise. Also shown are pitch angles of the CO arms (\emph{filled
squares with error bars} --- most errors are slightly smaller than the size of
the squares). The azimuthal angle is measured counter-clockwise from the top of
the face-on galaxy and the gas flow relative to the spiral pattern is from left
to right.}
\label{fig:co+psi} 
\end{figure*}

Figure~\ref{fig:co+psi} shows how the orientation of the magnetic field lines
varies with azimuthal angle at radii in the range $0.9\kpc<r<6.5\kpc$. The pitch
angles of the two CO spiral arms are also indicated (see
Sect.~\ref{sec:anisotropic}). These show the positions of the gaseous spiral
arms and let us measure how well the orientations of the magnetic field
\emph{lines} and spiral \emph{structure} agree.

The first thing to note is that the orientation of the magnetic field lines is
almost never zero. This means that the regular magnetic field has a spiral shape
--- and therefore both a radial and an azimuthal component --- virtually
everywhere in the region shown. The orientation of the regular magnetic field is
consistently less than $45\degr$ at the radii shown ($\pB>45\degr$ only for
small regions at $r>5.3\kpc$); the azimuthal component of the magnetic field
dominates the radial component for $r<6.5\kpc$.

Figure~\ref{fig:co+psi} also shows that the orientation of the regular magnetic
field lines vary by at least $\sim 30\degr$ in each ring; this is by no means
obvious in maps showing the B-vectors (e.g. Fig.~\ref{fig:orig_maps}d), where
the field orientation appears rather constant. The same situation may also exist
in observations of other external galaxies, whereby the field \emph{appears} to
be better aligned with the optical spiral and have a more consistent orientation
than is really the case. Note that even for galaxies with a small inclination to
the line of sight a slice at constant radius is quite elliptical and it is hard
to judge ``by sight'' how the orientation of the field varies with azimuth.

The other striking feature of Fig.~\ref{fig:co+psi} is the extremely good
agreement between the CO spiral arm pitch angles and the magnetic field
orientation at the position of the CO arms. This cannot be a coincidence as
$\pB$ itself varies considerably at each radius and the agreement holds whether
$\pco$ is low (e.g. $\pco\approx 20\degr$ at $r=1.6\kpc$) or high (e.g.
$\pco\approx 40\degr$ at $r=5.0\kpc$). Thus the orientation of the regular
magnetic field and the gaseous spiral arms are tightly linked in M51. There is
not a consistent trend in $\pB$ away from the spiral arms. The magnetic field
orientation is sometimes higher and sometimes lower in the inter-arm region than
at the position of the gaseous spiral arms. 

Berkhuijsen et al.~(\cite{Berkhuijsen97}) found a broad agreement between
optical arm pitch angles (Howard \& Byrd \cite{Howard90}), averaged in $\sim
5\kpc^2$ sectors, and their large scale magnetic field model of M51.

\begin{figure}[htbp] 
\includegraphics[width=0.45\textwidth]{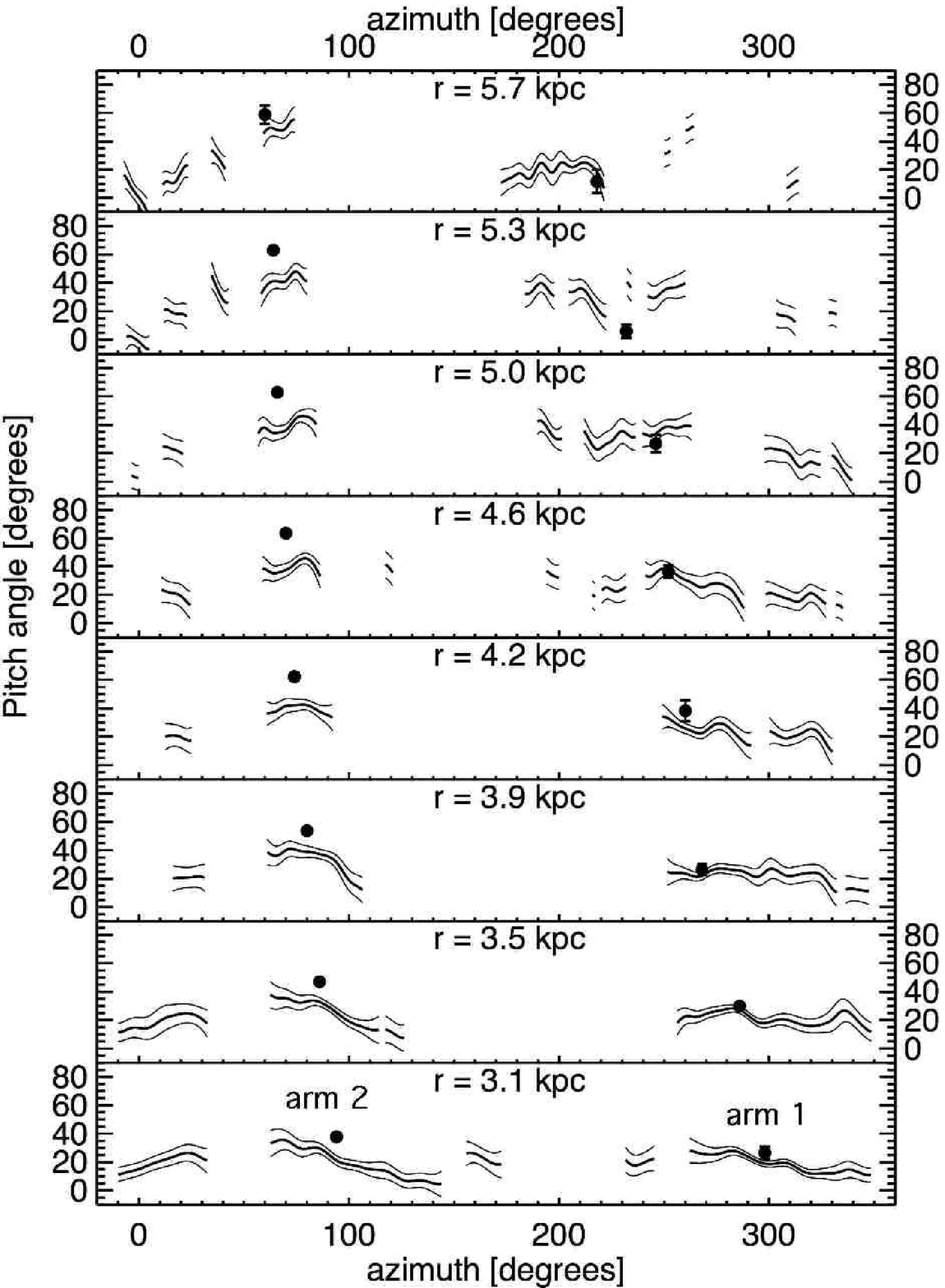}
\caption{As in Fig.~\ref{fig:co+psi}, lines show the orientation of the regular
magnetic field, but the \emph{filled circles} with error bars represent the
pitch angle of the ridges of \wav{6} polarized emission, $\ppi$. The polarized
ridge running from azimuth $\sim 100\degr$ at the bottom to $\sim 60\degr$ at
the top is the prominent inter-arm polarization ridge in
Fig.~\ref{fig:ridges}.} 
\label{fig:pi+psi} 
\end{figure}

In Fig.~\ref{fig:pi+psi} we compare $\pB$ with $\ppi$. In polarization arm~1
$\pB\simeq \ppi$; the regular magnetic field orientation is parallel to the
ridge of the PI6 arm. This indicates that the same physical effect is
responsible for both the increase in polarized emission along the ridge and the
orientation of the regular magnetic field there. This section of polarization
arm 1 lies $100$-$200\pc$ upstream of the corresponding arms in other tracers.
In contrast, along polarization arm~2, $\pB<\ppi$; the magnetic field always has
a lower pitch angle than the polarization arm, with a difference in angle of
$10$--$20\degr$. This section of polarization arm~2 lies about $\sim400\pc$
upstream of the other tracers of arm 2 and proceeds to cross over the gas and
optical spirals at a radius of $r\simeq 5.5\kpc$. This indicates that different
forces are at work in producing polarization arms 1 and 2.

\section{Discussion}
\label{sec:discuss}

New observations of the spiral galaxy M51, obtained with the VLA and Effelsberg
radio telescopes, and published maps of the molecular (CO) and mid-infrared
(15$\mu$m) emissions, were used to compare the spiral patterns present in the
magnetic field, dense gas and dust distributions.

1. We have shown that anisotropic wavelet functions are appropriate tools to
locate the positions of spiral arms in images of galaxies and to measure their
local pitch angles.

2. A systematic offset between the spiral ridges of molecular gas (CO) emission
and of mid-infrared (ISO 15$\mum$) emission can be followed over many kpc along
the arms of M51 (Figs.~\ref{fig:ridges} \& \ref{fig:log_ridges}). The typical
offset is $100\pc$ corresponding to a typical time delay of $\sim 10^{7}\yr$ for
gas well inside the co-rotation radius, assuming a circular rotation velocity of
$200\kms$ (Garcia-Burillo et al. \cite{GarciaBurillo93}), pattern speed of
$38\kms\kpc^{-1}$ (Zimmer et al. \cite{Zimmer04}) and arm inclination of
$20\degr$. The delay can be interpreted as the interval between the formation of
Giant Molecular Clouds (GMCs) and the appearance of newly formed stars. A delay
of $\sim 10^{7}\yr$ between dense gas accumulation and star formation was also
estimated by Tosaki et al. (\cite{Tosaki02}) from an analysis of the location of
$^{12}$CO and $^{13}$CO emission lines.

We do not see a shift in the relative positions of the ISO and CO arms,
i.e. the CO arm does not move from the concave to the convex side of the ISO arm,
near to the co-rotation radius of $r\simeq 6.2\kpc$ determined by
Elmegreen et al.~(\cite{Elmegreen92}). This may be due to the non-trivial dynamical
effects of two, overlapping, spiral arm patterns in M51 
(Elmegreen et al.~\cite{Elmegreen89}).

3. We also found a systematic offset between the spiral ridges of total radio
continuum emission and those of mid-infrared and CO emission, indicating a shift
between regions of strong thermal and non-thermal radio continuum emission.
While the thermal component is expected to closely follow the distribution of
star-forming regions as traced by mid-infrared emission, the non-thermal
component is probably enhanced due to compression of magnetic fields in a spiral
shock front, located upstream of the mid-infrared spiral arm. Cosmic ray
electron density is expected to be higher on the downstream side of the arms,
where supernova remnants should be more common, but on scales of a few hundred
parsecs this may not have a significant effect on the synchrotron intensity; we
are assuming that the distribution of cosmic ray electrons is mainly determined
by the magnetic field, rather than the distribution of their sources. The purely
non-thermal, polarized radio ridges lie the furthest upstream of all the data as
expected in this scenario. This is in agreement with the results of Tilanus et
al. (\cite{Tilanus88}) who found a general arm sequence of: non-thermal emission
-- (cold) dust lane -- thermal emission -- H$\alpha$.

Inside co-rotation our observations reveal the following sequence from upstream
to downstream: polarized radio emission, total radio emission, CO, infrared
emission, interpreted as a result of the sequence of components:
shock-compressed magnetic field, molecular gas clouds, and UV heated (warm)
dust. This can arise if the spiral shock is located slightly ahead of the CO
ridge, a plausible consideration if molecule formation and/or cloud condensation
are \emph{triggered} by the spiral shock. The compression of magnetic field,
resulting in stronger polarized radio emission, occurs at the shock, but the
formation of dense CO clouds is only completed a few hundred parsec downstream.
Following similar arguments to those used in point 1 above, we can estimate the
time delay between gas entering the shock front and the accumulation of dense
GMCs: a typical polarized ridge offset of $300\pc$ corresponds to an interval of
$\sim 10^{7}\yr$. The total time between gas entering the spiral shock and the
emergence of newly formed stars will then be a few tens of millions of years.

The above discussion only applies to a diffuse component of the CO in the
inter-arm region, which can be compressed by a large scale shock at the spiral
arm. Molecular gas that is already gravitationally bound in clouds will not
respond to the shock. Extended, faint CO(1-0) emission in the inner inter-arm
region of M51 is clearly seen in the combined single-dish and interferometer
data of Helfer et al.~(\cite{Helfer03}), suggesting that at least a part of the
molecular gas is diffuse. However the higher resolution, interferometer only,
CO(1-0) maps of Aalto et al.~(\cite{Aalto99}) reveal small clumps of inter-arm
molecular gas, so that some --- difficult to estimate --- fraction of the
upstream gas is probably already confined in clouds. This means that the shock
strength will be difficult to infer from gas density ratios in the inter-arm and
arm regions. On the other hand, Aalto et al. could identify steep velocity
gradients both in and, particularly clearly, between giant associations of
molecular gas which are in broad agreement with theoretical expectations of
spiral density wave shocks. The detailed physics of the gas response to possible
shocks in M51 and the effect this has on the spiral arm morphology requires more
sophisticated modelling than we can attempt here, especially the apparent
upstream shift of the shock in the magnetic field; there are, however, strong
indications that large scale spiral shocks occur in the vicinity of the gas arms
and that these compress diffuse gas.

4. The maintenance of both radial and azimuthal components of the magnetic
field, and hence a non-zero pitch angle, is an important prediction of dynamo
theory (Beck et al. \cite{Beck96}, Shukurov\ \cite{Shukurov00}) but is very
difficult to explain if the magnetic field is purely passive and stretched by
shear in the galactic differential rotation, since after a small number of
galactic rotations the field would be completely circular, i.e. it would have a
pitch angle $\pB\simeq 0$.

5. Although the B-vectors of polarized radio emission from M51 seem to have the
same orientations as the CO arms (Fig.~\ref{fig:orig_maps}), they smoothly
change by about $\pm15\degr$ around any fixed radius, without a systematic trend
between the arm and inter-arm regions (Fig.~\ref{fig:co+psi}). Smaller pitch
angles in inter-arm regions are predicted by some dynamo models and have been
observed e.g. in NGC~6946 (Rohde et al.~\cite{Rohde99}). In M51, however, the
regular magnetic field structure is probably dominated by streaming motions and
compression in the spiral shock (see below). The field is brought into good
alignment at the CO arms \emph{or vice versa} (Fig.~\ref{fig:co+psi}) and then
relaxes into a different configuration in the inter-arm regions that depends on
e.g. the amount of compression in the arms, the energy of turbulent flows, the
orientation of dynamo-generated field modes etc.

6. We found a close alignment of the CO pitch angles and the regular magnetic
field orientation \emph{at the position of the arms} (Fig.~\ref{fig:co+psi}).
This could be due to shock amplification of the regular field component parallel
to the shock. 

If the field is frozen into the gas then
$\Brpa^{(s)}=\epsilon\Brpa$ and $\Brpe^{(s)}=\Brpe$, where
$\epsilon=\rho^{(s)}/\rho$ is the compression ratio in the gas density $\rho$
and the superscript $(s)$ labels quantities at the shock front. The deflection
of the magnetic field $\Delta\theta$, where $\theta$ is the angle between the
field and the shock front upstream of the shock front, is given by
\begin{eqnarray} 
\label{eq:align}
\Delta\theta&=&\theta-\theta^{(s)}=\arctan\left(\frac{\Brpe}{\Brpa}\right)-
	\arctan\left(\frac{\Brpe}{\epsilon\Brpa}\right) \nonumber \\ 
	& \simeq & \frac{\Brpe}{\Brpa}\left(1-\frac{1}{\epsilon}\right) 
\end{eqnarray} 
where the last line holds for
$\Brpe/\Brpa\ll 1$. For $\epsilon=4$ (maximum degree of gas compression by an
adiabatic shock) and $\Brpe/\Brpa=0.5$ (i.e. the magnetic field is initially
inclined at $30\degr$ to the shock front, a conservative estimate c.f.
Fig.~\ref{fig:co+psi}) we obtain $\Delta\theta\simeq 20\degr$. Thus a strong
spiral shock can be expected to align the regular magnetic field rather well
with the spiral arm.

If the shock is weak ($\epsilon<4$) or the upstream field makes a larger angle
with the shock front then an additional source of alignment is required. The
conversion of an isotropic random magnetic field component upstream of the shock
into an anisotropic random field in the shocked region (where the anisotropic
random field will be perfectly aligned with the shock) provides such a
mechanism. Strong polarized emission arising from anisotropic random magnetic
fields formed in shearing shocks has recently been identified in the barred
galaxies NGC~1097 and NGC~1365 (Beck et al.~\cite{Beck05a}). 

7. Some of the polarized radio emission (and hence some of the regular magnetic
fields) forms ridges which are interlaced with the gaseous arms: the polarized
ridge of arm~2 clearly lies in between gas arms~1 and 2 south-east of the
central region. The magnetic field is \emph{not} oriented parallel to this
polarization arm but is inclined to the axis of the arm by up to $20\degr$
(Fig.~\ref{fig:pi+psi}). In contrast to polarization arm~1, spiral shock
compression does not seem to be the origin of these enhanced regular magnetic
fields. The field orientation seems to be between that of the polarization arm
and that of the nearby CO arm~2, like finding a compromise between two different
forces. One force causes the polarization arm to cross CO arm~2 to join the
outer arm~1, the second tries to align the field with the gas arm.

\section*{Acknowledgements} 
We thank Anvar Shukurov for useful discussions and the referee for helpful
comments. This work was supported in part by the DFG-RFBR (grant 03-02-04031).
I.P. is grateful to the Max-Planck-Institut f\"ur Radioastronomie for support
and hospitality. A.F. acknowledges the Leverhulme Trust for financial support
under research grant F/00 125/N. This research made use of NASA's Extragalactic
Database (NED) and Astrophysical Data System (ADS).


\end{document}